\documentclass[apj]{emulateapj}
\usepackage{mathptmx}
\usepackage{lscape}
\usepackage{rotate}

\newcommand{\aox}{$\alpha_{\rm ox}$}
\newcommand{\daox}{$\Delta\alpha_{\rm ox}$}
\newcommand{\chandra}{{\it Chandra\/}}
\newcommand{\xmm}{{\it XMM-Newton\/}}
\newcommand{\rosat}{{\it ROSAT\/}}
\newcommand{\xray}{\hbox{X-ray}}
\def\gtsim{\lower 2pt \hbox{$\, \buildrel {\scriptstyle >}\over  {\scriptstyle \sim}\,$}}
\def\ltsim{\lower 2pt \hbox{$\, \buildrel {\scriptstyle <}\over  {\scriptstyle \sim}\,$}}
\newcommand{\luv}{\hbox{$L_{\mbox{\scriptsize 2500~\AA}}$}}
\newcommand{\lx}{\hbox{$L_{\rm 2~keV}$}}

\journalinfo{The Astrophysical Journal, 666:??--??, 2007 September 10}
\slugcomment{Received 2007 February 13; accepted 2007 May 18}

\shorttitle{X-RAY STUDY OF THE MOST-LUMINOUS QUASARS}
\shortauthors{JUST ET AL.}

\begin{document}

\title{The X-ray Properties of the Most-Luminous Quasars from the \\
Sloan Digital Sky Survey}

\author{D.~W.~Just\altaffilmark{1},
        W.~N.~Brandt\altaffilmark{1},
        O.~Shemmer\altaffilmark{1},
        A.~T.~Steffen\altaffilmark{1},
        D.~P.~Schneider\altaffilmark{1},
        G.~Chartas\altaffilmark{1},
        and G.~P.~Garmire\altaffilmark{1}
	}
\altaffiltext{1}{Department of Astronomy and Astrophysics, 525 Davey
Laboratory, The Pennsylvania State University, University Park, PA
16802; dwjust@astro.psu.edu}

\begin{abstract}

Utilizing 21 new \chandra\ observations as well as archival \chandra,
\rosat, and \xmm\ data, we study the \xray\ properties of a
representative sample of 59 of the most optically luminous quasars in
the Universe ($M_i\approx -29.3$ to $-30.2$) spanning a redshift range
of $z\approx1.5$--4.5. Our full sample consists of 32 quasars from the
Sloan Digital Sky Survey (SDSS) Data Release 3 (DR3) quasar catalog,
two additional objects in the DR3 area that were missed by the SDSS
selection criteria, and 25 comparably luminous quasars at
$z\gtsim4$. This is the largest \xray\ study of such luminous quasars
to date. By jointly fitting the \xray\ spectra of our sample quasars,
excluding radio-loud and broad absorption line (BAL) objects, we find
a mean \xray\ power-law photon index of
\hbox{$\Gamma=1.92^{+0.09}_{-0.08}$} and constrain any neutral
intrinsic absorbing material to have a mean column density of $N_{\rm
H}$~\ltsim~$2 \times 10^{21}$~${\rm cm}^{-2}$. We find, consistent
with other studies, that $\Gamma$ does not change with redshift, and
we constrain the amount of allowed $\Gamma$ evolution for the
most-luminous quasars. Our sample, excluding radio-loud and BAL
quasars, has a mean \xray-to-optical spectral slope of
\hbox{\aox~$=-1.80\pm0.02$}, as well as no significant evolution of
\aox\ with redshift. We also comment upon the \xray\ properties of a
number of notable quasars, including an \xray\ weak quasar with
several strong narrow absorption-line systems, a mildly radio-loud BAL
quasar, and a well-studied gravitationally lensed quasar.

\end{abstract}


\section{Introduction}

The most optically luminous known quasars ($M_i<-29$) serve as
valuable astrophysical probes of extreme accretion conditions and the
distant universe. These objects have been found to date at
\hbox{$z\approx 1.5$--4.5}; their resulting $i$ magnitudes of
\hbox{$\approx 15$--18} and relatively bright multiwavelength fluxes
allow them to be studied effectively with a variety of facilities
across the electromagnetic spectrum. Even if they are radiating near
the Eddington limit with $L/L_{\rm Edd}\approx 1$, their energy
outputs require \hbox{$\approx 10^9$--$10^{10}$~M$_\odot$} nuclear
black holes and thus they are presumably associated with the
most-massive galaxies; today many of these objects have likely evolved
into supergiant ellipticals found in the cores of rich clusters. As
the most-luminous, non-transient emitters at high redshift, these
quasars have been useful in cosmological studies including measuring
absorption lines from intervening line-of-sight material (e.g., Rauch
1998; Wolfe et~al.\ 2005; and references therein), assessing the Cold
Dark Matter cosmogony (e.g., Efstathiou \& Rees 1988; Turner 1991;
Springel et~al.\ 2005), and constraining the accretion history of the
Universe (e.g., Croom et~al.\ 2004; Richards et~al.\ 2006).

The Sloan Digital Sky Survey (SDSS; York et~al.\ 2000) is now
providing the most-complete selection of highly luminous quasars to
date (e.g., Schneider et~al.\ 2005, hereafter S05). A large fraction
of these objects, about 2/3, lack pointed or serendipitous \xray\
detections (aside from at $z>4$, where pointed \xray\ observations
have detected a large fraction; e.g., Vignali et~al.\ 2003,
2005). Accordingly, we have started a project aimed at improving
understanding of the \xray\ properties of the most-luminous known
quasars over as broad a redshift range as possible.  The \hbox{X-ray}
emission from quasars probes the innermost regions of their
accretion-disk coronae where any changes in the mode of accretion
might be most evident, and \hbox{X-ray} spectroscopy provides
constraints on intrinsic and intervening absorption.

Recent studies of the \xray\ spectra (e.g., Page et~al.\ 2005; Shemmer
et~al.\ 2005a, 2006a; Vignali et~al.\ 2005) and X-ray-to-optical
spectral energy distributions (SEDs; e.g., Strateva et~al.\ 2005;
Steffen et~al.\ 2006, hereafter S06) of quasars have generally shown
no clear changes with redshift, although some exceptions have been
found (e.g., Grupe et~al.\ 2006; Kelly et~al.\ 2007) and at lower
luminosities \xray\ spectral evolution may be observed (Dai et~al.\
2004). These results indicate that the inner regions of quasars are
largely insensitive to the enormous changes in large-scale cosmic
environment occurring over the history of the Universe. There is
evidence, however, that the photon index ($\Gamma$) of the \xray\
power-law spectrum increases as $L/L_{\rm Edd}$ increases (e.g.,
Shemmer et~al.\ 2006b) and that the X-ray-to-optical flux ratio (\aox;
Tananbaum et~al.\ 1979) drops with increasing luminosity. Studies of
the luminosity and redshift dependence of $\Gamma$ and \aox\ benefit
from the widest possible sampling of the luminosity-redshift plane;
such wide coverage is needed to break the luminosity-redshift
degeneracy invariably present in flux-limited samples. By
systematically studying the most-luminous quasars over the full
redshift range where they exist in the Universe, \hbox{$z\approx
1.5$--4.5}, it is possible to populate one important region of this
plane, complementing efforts to fill other regions of
luminosity-redshift space (e.g., S06). Systematic \xray\ measurements
of the most-luminous quasars also serve to broaden the well-sampled
luminosity range available for study and thereby minimize the
possibility of confusion by spurious correlations (e.g., Yuan et~al.\
1998).

In this paper we study, using a combination of new \chandra\
``snapshot'' observations as well as archival \chandra, \rosat, and
\xmm\ data, the basic \xray\ properties of 32 of the 33 most-luminous
quasars in the SDSS Data Release 3 (DR3) quasar catalog (S05; see
\S2.1.1 for a discussion of the one quasar that is not included in our
study). All 32 of the quasars in our SDSS sample have \xray\
detections. We also include two comparably luminous quasars missed by
the SDSS selection and an additional 25 comparably luminous non-DR3
quasars at $z\gtsim4$. We use our results to strengthen constraints
upon the X-ray spectral and \xray-to-optical SED properties of the
most-luminous quasars, via a combination of single-object and
multiple-object analyses.

We detail the general properties of our sample in \S2, as well as the
\xray\ observations and data reduction. \xray, optical, and radio
properties are presented in \S3, and optical spectra and notes on
exceptional objects appear in \S4. Data analysis and results are given
in \S5, and a summary of our findings is given in \S6. We adopt a
cosmology with $H_0=70$~km~s$^{-1}$~Mpc$^{-1}$, $\Omega_{\rm M}=0.3$,
and $\Omega_\Lambda=0.7$.

\section{Sample and X-ray Data}

\subsection{Sample Selection and Properties}

\subsubsection{SDSS DR3 Quasars}

About half of our sample of highly luminous quasars has been drawn
from the SDSS DR3 quasar catalog (S05). The SDSS, an optical imaging
and spectroscopic survey that aims to cover about one-quarter of the
entire sky, targets active galaxies for follow-up spectroscopy
primarily based upon their $ugriz$ (Fukugita et~al.\ 1996) colors and
magnitudes (e.g., Richards et~al.\ 2002). Active-galaxy candidates at
$z\ltsim3$ are spectroscopically targeted if their $i$ magnitudes are
\hbox{$15$--19.1}; high-redshift candidates are targeted if
\hbox{$i=15$--20.2} (the limit at $i=15$ is imposed to avoid
saturation and fiber cross-talk problems in the SDSS spectroscopic
observations). The DR3 quasar catalog has been constructed from SDSS
spectroscopic observations over a solid angle of 4188 deg$^2$ (about
10\% of the sky). Given the large areal coverage, this catalog should
contain representative members of the population of the most
optically luminous quasars in the Universe; i.e., other surveys are
unlikely to find a population of quasars significantly more luminous
than those studied here.\footnote{The recently released SDSS DR5
quasar catalog (Schneider et~al.\ 2007), covering 5740 deg$^2$,
further supports this assertion.} About 60\% of the most
optically luminous quasars in the SDSS DR3 quasar catalog had been
discovered in earlier surveys, such as the Hamburg Quasar Survey
(e.g., Hagen et~al.\ 1999), the Second Byurakan Survey (e.g.,
Stepanian et~al.\ 2001), the University of Michigan Survey (e.g.,
MacAlpine \& Lewis 1978), and the Palomar Digital Sky Survey (DPOSS;
e.g., Djorgovski et~al.\ 1998).

\begin{figure}
%\epsscale{0.8}
\plotone{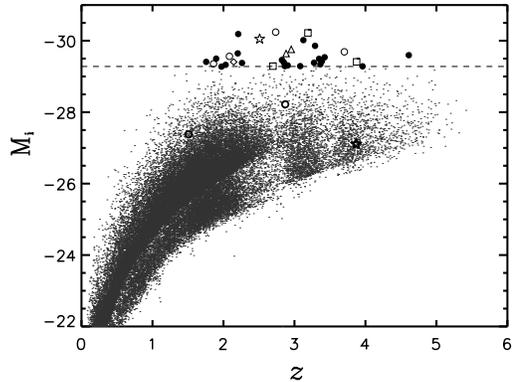} 
\caption{\label{dr3_mi_plot} Absolute $i$-band magnitude vs. redshift
 for our SDSS sample compared with the SDSS DR3 quasar catalog. Our
 SDSS sample of 32 quasars includes both archival (open symbols) and
 targeted (filled symbols) sources with \chandra\ (circles), \xmm\
 (triangles), and \rosat\ (squares) observations. The two additional
 sources that were missed by the SDSS (APM~08279$+$5255 and
 HS~1603$+$3820; see \S2.1.2) are shown as stars. The gravitationally
 lensed quasars APM~08279$+$5255, SDSS~J0145$-$0945, and
 SDSS~J0813$+$2545 have been de-amplified to their true $M_i$
 magnitudes and have bold symbols for clarity; all fail our cutoff at
 $M_i<-29.28$, which is shown as a dashed line. SDSS~J1007+0532, which
 has been targeted by S.~F. Anderson and thus needed to be removed
 from our sample (see \S2.1.1), is shown as an open diamond. Small
 dots represent the $\approx46,000$ quasars in the DR3 catalog.  }
\end{figure}

We sorted the DR3 quasar catalog upon $M_i$ and considered the 33
most-luminous quasars in the catalog for \chandra\ targeting (see
Figure~\ref{dr3_mi_plot}). The number 33 was chosen based upon
practical \xray\ observing-time considerations, and this sample size
is large enough to provide statistically meaningful results. Of the 33
most-luminous quasars, 11 already had detections in archival \xray\
data and were not targeted; these archival data have been utilized in
our study. The remaining 22 quasars were proposed via the \chandra\
Cycle~7 Guaranteed Time Observing program, and 21 of them were awarded
observing time. One of our targets, SDSS~J100711.81$+$053208.9, was
awarded to another \chandra\ observer (S.F. Anderson) as part of a
program studying bright and extreme Broad Absorption Line (BAL)
quasars.  We do not consider the omission of SDSS~J1007$+$0532 from
our sample to be statistically problematic. In fact, owing to its
BAL-quasar nature, this object would need to be removed from most of
our analyses of \aox, $\Gamma$, and other properties in any case. Our
SDSS sample thus includes 32 quasars with $M_i$ values of $-29.28$ to
$-30.24$, all of which have sensitive \xray\ coverage; we adopt
$M_i=-29.28$ as a practical minimum luminosity for our sample. These
32 quasars span essentially the entire range of redshift
(\hbox{$z\approx 1.5$--4.5}) over which such luminous objects are
known, although the source statistics at $z>4$ are limited.

\subsubsection{Incompleteness and Complementary z\gtsim4 Quasars}

For the highly luminous and optically bright broad-line quasars under
consideration here, the SDSS is not expected to suffer from
substantial incompleteness biases. However, some incompleteness is
expected due to the SDSS spectroscopic limit of $i=15$ and the fact
that, at \hbox{$z\approx 2.5$--2.9} and \hbox{$z\approx 3.4$--3.6},
the SDSS colors of quasars intersect the stellar locus (e.g., Richards
et~al.\ 2002, 2006; S05). Furthermore, about 5\% of quasars are
expected to be missed by the SDSS, largely due to image defects and
source blending (Vanden Berk et~al.\ 2005). We have searched for
missed quasars more luminous than $M_i=-29.28$ in the area covered by
the DR3 quasar catalog using the NASA Extragalactic
Database\footnote{See http://nedwww.ipac.caltech.edu/.} (NED) combined
with accurate SDSS photometry (since the photometry in NED is not
uniform and can contain significant errors). We have found only two
missed quasars: APM~08279+5255 at $z=3.91$ (e.g., Irwin et~al.\ 1998)
and HS~1603+3820 at $z=2.51$ (e.g., Dobrzycki et~al.\
1999).\footnote{Our NED search also uncovered the object
QUEST~J150724.0$-$020212.8 in the SDSS area, which is identified as an
exceptionally luminous $z=1.09$ quasar by Rengstorf et~al.\
(2004). However, that identification relied on the assumption that the
single feature seen in a spectrum was the Mg~{\sc ii} line. Given this
tenuous identification we obtained a spectrum of this source using the
Low Resolution Spectrograph (LRS) on the Hobby-Eberly Telescope (HET;
Ramsey et~al.\ 1998). Our spectrum does not show the emission feature
seen by Rengstorf et~al., and we identify this object as a Galactic
star and not a quasar.}  APM~08279+5255 is a gravitationally lensed
BAL quasar with $i=14.9$ that slightly violated the SDSS spectroscopic
limit. HS~1603+3820 has a very rich spectrum of associated heavy
element absorbers (e.g., Misawa et~al.\ 2005); it is near the redshift
range where the SDSS colors of quasars intersect the stellar locus and
was missed by the SDSS quasar color-selection algorithm by just
0.02~mag in $u$ (G.~T.~Richards 2006, private
communication). Fortunately, both APM~08279+5255 (Chartas et~al.\
2002) and HS~1603+3820 (PI: Dobrzycki) have \xray\ detections in
archival pointed \chandra\ observations, and we include these quasars
in our analyses below as appropriate. After these two additions, we
expect $\ltsim15$\% incompleteness for $M_i<-29.28$ broad-line quasars
at \hbox{$z\approx1.5$--4.5} in the area covered by the SDSS DR3
quasar catalog. The observation log of our sample, including these two
additional sources, appears in Table~1; this 34 object sample will
hereafter be referred to as the ``core'' sample.

\begin{deluxetable*}{lclccll}
\tablecolumns{7} \tablenum{1} \tabletypesize{\tiny}
\tablecaption{X-ray Observation Log of the Core Sample}
\tablehead{
\colhead{} & \colhead{} & \colhead{X-ray} & \colhead{Chandra} &
\colhead{Exp. Time\tablenotemark{a}} & \colhead{Alternate} &
\colhead{} \\ \colhead{Object (SDSS J)} & \colhead{$z$} &
\colhead{Obs. Date} & \colhead{Cycle No.}  & \colhead{(ks)} &
\colhead{Designation} & \colhead{Notes\tablenotemark{b}} } \startdata
\dataset[ADS/Sa.CXO#Obs/6827]{012156.04$+$144823.9} & 2.87 & 2006 Jun
21 & 7 & 3.9 & HS 0119+1432 & PC \\
\dataset[ADS/Sa.CXO#Obs/5192]{014516.59$-$094517.3} & 2.73 & 2004 Aug
23 & 5 & 13.3 & UM 673 & AC, Lensed \\
\dataset[ADS/Sa.CXO#Obs/6820]{020950.71$-$000506.4} & 2.85 & 2005 Dec
02 & 7 & 2.2 & UM 402 & PC \\
\dataset[ADS/Sa.CXO#Obs/6829]{073502.31$+$265911.4} & 1.97 & 2005 Dec
05 & 7 & 4.0 & \nodata & PC \\
\dataset[ADS/Sa.CXO#Obs/6814]{075054.64$+$425219.2} & 1.90 & 2006 Sep
05 & 7 & 4.0 & HS 0747+4259 & PC \\
\dataset[ADS/Sa.CXO#Obs/6824]{080342.04$+$302254.6} & 2.03 & 2006 Sep
06 & 7 & 4.0 & HS 0800+3031 & PC \\
\dataset[ADS/Sa.CXO#Obs/3023]{081331.28$+$254503.0} & 1.51 & 2002 Jan
30 & 3 & 5.0 & HS 0810+2554 & AC, Lensed \\
\dataset[ADS/Sa.CXO#Obs/6815]{084401.95$+$050357.9} & 3.35 & 2006 Feb
18 & 7 & 3.8 & \nodata & PC, BAL, RLQ \\
\dataset[ADS/Sa.CXO#Obs/6810]{090033.49$+$421546.8} & 3.29 & 2006 Feb
09 & 7 & 3.9 & HS 0857+4227 & PC \\
\dataset[ADS/Sa.CXO#Obs/6821]{094202.04$+$042244.5} & 3.28 & 2006 Feb
08 & 7 & 4.1 & \nodata & PC \\
\dataset[ADS/Sa.CXO#Obs/6828]{095014.05$+$580136.5} & 3.96 & 2006 Jan
29 & 7 & 3.9 & PSS J0950+5801 & PC \\
\dataset[ADS/Sa.CXO#Obs/6819]{100129.64$+$545438.0} & 1.76 & 2006 Jan
29 & 7 & 4.0 & Mrk 132 & PC \\
\dataset[ADS/Sa.CXO#Obs/6809]{101447.18$+$430030.1} & 3.13 & 2006 Jun
14 & 7 & 4.1 & HS 1011+4315 & PC \\
\dataset[ADS/Sa.CXO#Obs/6811]{110610.73$+$640009.6} & 2.20 & 2006 Jul
16 & 7 & 3.7 & HS 1103+6416 & PC \\ 111038.64$+$483115.6 & 2.96 & 2002
Apr 25 & \nodata & 14.1 & Q1107+487 & AX, Page et al. (2005) \\
121930.77$+$494052.3 & 2.70 & 1992 May 07 & \nodata &
2.2\tablenotemark{c} & SBS 1217+499 & AR \\
\dataset[ADS/Sa.CXO#Obs/6816]{123549.47$+$591027.0} & 2.82 & 2006 Jul
29 & 7 & 3.9 & SBS 1233+594 & PC \\
\dataset[ADS/Sa.CXO#Obs/6817]{123641.46$+$655442.0} & 3.39 & 2006 Aug
29 & 7 & 4.0 & \nodata & PC \\
\dataset[ADS/Sa.CXO#Obs/6825]{135044.67$+$571642.8} & 2.91 & 2006 Oct
15 & 7 & 3.0 & SBS 1348+575 & PC \\
\dataset[ADS/Sa.CXO#Obs/6826]{140747.22$+$645419.9} & 3.08 & 2006 Sep
16 & 7 & 3.7 & \nodata & PC \\
\dataset[ADS/Sa.CXO#Obs/6823]{142123.98$+$463317.8} & 3.37 & 2006 Sep
15 & 7 & 3.9 & \nodata & PC \\ 142656.17$+$602550.8 & 3.19 & 1993 Nov
01 & \nodata & 4.1\tablenotemark{c} &SBS 1425+606 & AR, Reimers et
al. (1995) \\ \dataset[ADS/Sa.CXO#Obs/6812]{143835.95$+$431459.2} &
4.61 & 2006 Sep 16 & 7 & 3.7 & \nodata & PC \\ 144542.75$+$490248.9 &
3.88 & 1993 Jul 12 & \nodata & 5.7\tablenotemark{c} & & AR \\
\dataset[ADS/Sa.CXO#Obs/6808]{152156.48$+$520238.4} & 2.19 & 2006 Jul
16 & 7& 4.1 & \nodata & PC \\ 152553.89$+$513649.1 & 2.88 & 2001 Dec
08 & \nodata & 24.4 & CSO 755 & AX, BAL, Shemmer et al. (2005b) \\
\dataset[ADS/Sa.CXO#Obs/4127]{161434.67$+$470420.0} & 1.86 & 2003 Dec
20 & 4 & 2.6\tablenotemark{c} &RX J1614.5+4704 & AC, Bade et
al. (1995) \\ \dataset[ADS/Sa.CXO#Obs/2184]{162116.92$-$004250.8} &
3.70 & 2001 Sep 05 & 2 & 1.6 & \nodata & AC, Bechtold et al. (2003) \\
\dataset[ADS/Sa.CXO#Obs/547]{170100.62$+$641209.0} & 2.74 & 2000 Oct
31 & 1 & 39.4\tablenotemark{c} &HS 1700+6416 & AC, Reimers et
al. (1995) \\ \dataset[ADS/Sa.CXO#Obs/6813]{173352.22$+$540030.5} &
3.43 & 2006 May 28 & 7 & 3.7 & \nodata & PC, RLQ \\
\dataset[ADS/Sa.CXO#Obs/6822]{212329.46$-$005052.9} & 2.26 & 2006 Mar
30 & 7 & 3.9 & \nodata & PC \\
\dataset[ADS/Sa.CXO#Obs/5701]{231324.45$+$003444.5} & 2.08 & 2005 Sep
23 & 6 & 1.1 & Hazard 2310+0018 & AC, BAL \\ \cline{1-7}
\dataset[ADS/Sa.CXO#Obs/2979]{APM 08279$+$5255} & 3.91 & 2002 Feb 24 &
3 & 88.8 & \nodata & AC, BAL, Lensed, Chartas et al. (2002) \\
\dataset[ADS/Sa.CXO#Obs/4026]{HS 1603$+$3820} & 2.51 & 2002 Nov 29 & 4
& 8.3 & \nodata & AC \\ \enddata \tablenotetext{a}{The \chandra\
exposure time has been corrected for detector dead time.}
\tablenotetext{b}{PC---pointed \chandra\ observation; AC---archival
\chandra\ observation; AR---archival \rosat\ observation;
AX---archival \xmm\ observation. For sources with archival \xray\
data, we list the paper where the data were originally published, when
possible.}  \tablenotetext{c}{The mean effective exposure time for the
aperture used for sources at large off-axis angles.}
\end{deluxetable*}

In some analyses below, we will complement our core sample with 25
additional comparably luminous ($M_i$\ltsim$-29$) quasars at
$z\gtsim4$ that lie outside the area covered by the SDSS DR3 quasar
catalog. The \xray\ properties of these quasars have been studied by
Vignali et~al.\ (2003, 2005) and Shemmer et~al.\ (2005a) and appear in
Table~2. These quasars were discovered by the DPOSS and other
large-area surveys, and their basic optical properties are consistent
with those of our SDSS quasars. This complementary sample is required
to improve our statistical coverage at $z>4$, which would otherwise be
poor. Since the \xray\ and optical properties of these objects were
investigated as a whole in S06, this subsample will be referred to as
the complementary high-luminosity $z\gtsim4$ quasars from S06. We have
not added highly luminous quasars outside the SDSS DR3 area at $z<4$
from other surveys, since such quasars generally do not have
systematic sensitive \xray\ coverage.

\begin{deluxetable*}{cccccc}
\tablecolumns{6} \tablenum{2} \tabletypesize{\tiny}
\tablecaption{X-ray Properties of Complementary $z\gtsim4$ Sources
with $M_i\ltsim-29$} \tablehead{ \colhead{} & \colhead{} &
\colhead{Total \xray} & \colhead{} & \colhead{} & \colhead{} \\
\colhead{Name} & \colhead{$z$} & \colhead{Counts} & \colhead{$f_{\rm
x}$\tablenotemark{a}} & \colhead{\aox} &
\colhead{Reference\tablenotemark{b}} } \startdata
\dataset[ADS/Sa.CXO#Obs/3152]{PSS~0133$+$0400} & 4.15 & 36 & 20.8 &
$-$1.64 & 1 \\ \dataset[ADS/Sa.CXO#Obs/3018]{PSS~0134$+$3307} & 4.53 &
16 & 11.5 & $-$1.68 & 1 \\
\dataset[ADS/Sa.CXO#Obs/3153]{PSS~0209$+$0517} & 4.14 & 22 & 15.2 &
$-$1.75 & 1 \\ \dataset[ADS/Sa.CXO#Obs/876]{PSS~0248$+$1802} & 4.43 &
14 & 21.2 & $-$1.64 & 1 \\
\dataset[ADS/Sa.CXO#Obs/3156]{PSS~0955$+$5940} & 4.34 & 10 & 6.0 &
$-$1.81 & 1 \\ \dataset[ADS/Sa.CXO#Obs/3157]{PSS~0957$+$3308} & 4.20 &
17 & 11.3 & $-$1.74 & 1 \\
\dataset[ADS/Sa.CXO#Obs/878]{PSS~1057$+$4555} & 4.12 & 24 & 25.3 &
$-$1.70 & 1 \\ \dataset[ADS/Sa.CXO#Obs/3159]{PSS~1347$+$4956} & 4.51 &
30 & 17.5 & $-$1.78 & 1 \\
\dataset[ADS/Sa.CXO#Obs/875]{BR~0241$-$0146} & 4.06 & 12 & 4.3 &
$-$1.87 & 1 \\ \dataset[ADS/Sa.CXO#Obs/3031]{BR~0305$-$4957} & 4.73 &
3 & 2.1 & $-$1.94 & 1 \\
\dataset[ADS/Sa.CXO#Obs/3154]{BR~0311$-$1722}\tablenotemark{c} & 4.00
& 7 & 4.1 & $-$1.93 & 1 \\
\dataset[ADS/Sa.CXO#Obs/4064]{BR~0331$-$1622} & 4.36 & 14 & 8.5 &
$-$1.86 & 2 \\ \dataset[ADS/Sa.CXO#Obs/4065]{BR~0353$-$3820} & 4.55 &
55 & 44.7 & $-$1.54 & 2 \\
\dataset[ADS/Sa.CXO#Obs/4066]{BR~0418$-$5723} & 4.46 & 7 & 7.0 &
$-$1.87 & 2 \\ \dataset[ADS/Sa.CXO#Obs/4067]{BR~0424$-$2209} & 4.32 &
11 & 8.6 & $-$1.81 & 2 \\
\dataset[ADS/Sa.CXO#Obs/4068]{PSS~0747$+$4434} & 4.43 & 5 & 5.7 &
$-$1.81 & 2 \\ \dataset[ADS/Sa.CXO#Obs/4069]{PSS~1058$+$1245} & 4.33 &
5 & 2.4 & $-$2.07 & 2 \\ \dataset[ADS/Sa.CXO#Obs/4070]{BR~1117$-$1329}
& 3.96 & 2 & 1.8 & $-$2.03 & 2 \\
\dataset[ADS/Sa.CXO#Obs/4072]{PSS~1646$+$5514} & 4.04 & 5 & 5.0 &
$-$2.00 & 2 \\ \dataset[ADS/Sa.CXO#Obs/4073]{BR~2213$-$6729} & 4.47 &
20 & 14.9 & $-$1.64 & 2 \\
\dataset[ADS/Sa.CXO#Obs/4074]{PSS~2344$+$0342} & 4.24 &\nodata& $<2.6$
& $<-$1.98 & 2 \\ Q~0000$-$263 & 4.10 & 1229\tablenotemark{d} & 12.6 &
$-$1.70 & 3 \\ PSS~0926$+$3055 & 4.19 & 1156\tablenotemark{d} & 39.0 &
$-$1.76 & 3 \\ PSS~1326$+$0743 & 4.09 & 963\tablenotemark{d} & 27.9 &
$-$1.76 & 3 \\ BR~2237$-$0607 & 4.56 & 306\tablenotemark{d} & 8.5 &
$-$1.74 & 3 \\ \tableline \enddata \tablecomments{Quasars are first
sorted by reference, then further sorted by right ascension.}
\tablenotetext{a}{Galactic absorption-corrected flux in the observed
\hbox{0.5--2~keV} band in units of $10^{-15}$ erg cm$^{-2}$ s$^{-1}$.}
\tablenotetext{b}{(1) Vignali et~al.\ 2003; (2) Vignali et~al.\ 2005;
(3) Shemmer et~al. 2005a.}  \tablenotetext{c}{This source appears in
Vignali et~al.\ (2003) under its older designation BR~0308$-$1734.}
\tablenotetext{d}{Since these sources were observed by \xmm, we quote
the total counts from the pn detector.}
\end{deluxetable*}

\subsubsection{Radio-Loud, Broad Absorption Line, and Lensed Quasars}

Radio-loud quasars (RLQs) are known to have jet-linked \xray\ emission
components that generally lead to higher X-ray-to-optical luminosity
ratios than those of radio-quiet quasars (RQQs; e.g., Worrall et~al.\
1987). Given this finding, it is important to consider RLQs and RQQs
separately in statistical analyses of quasar \xray\ properties. We
quantify radio loudness using the radio-loudness parameter, {\it R\/},
defined as $R = f_{5\rm GHz}/f_{\mbox{\scriptsize 4400\AA}}$
(Kellermann et~al.\ 1989). We classify any quasar with $R>10$ as
radio-loud; details of our $R$ calculations are given in \S3. Two of
the quasars in our 34 object core sample, SDSS~J0844$+$0503 and
SDSS~J1733$+$5400, are RLQs. Both objects are only mildly radio loud
($R\approx19$ and $R\approx10$, respectively) and were observed as
part of our \chandra\ Cycle~7 observations. Given the radio coverage
of all of our sources (including tight upper limits of $R<0.5$--4 on
many of them), we do not expect there to be any RLQs that have not
been identified as such. Although it is possible that there may still
be some jet-linked \xray\ contribution for sources with $R<10$, we use
equation (2) of Worrall et~al.\ (1987) to estimate a limit of $\ltsim
2\%$ on the fraction of \xray\ emission from our RQQs that is jet
related.

Owing to intrinsic \xray\ absorption, BAL quasars also require special
consideration in statistical analyses of quasar \xray\ properties
(e.g., Gallagher et~al.\ 2006 and references therein). Using a catalog
of BAL quasars from DR3 compiled by Trump et~al.\ (2006), we found six
potential BAL quasars in our SDSS sample. The three we admit as BAL
quasars are SDSS~J0844$+$0503, SDSS~J1525$+$5136, and
SDSS~J2313$+$0034. All three have positive ``balnicity'' indices (see
Trump et~al.\ 2006); note that SDSS J0844$+$0503 is a mildly
radio-loud BAL quasar. The other three potential BAL quasars,
SDSS~J1001$+$5454, SDSS~J1407$+$6454, and SDSS~J1426$+$6025, have UV
absorption but do not formally satisfy the positive balnicity
criterion, so we do not remove them from our analyses below. The
removal of these three sources from our main analyses does not
significantly affect any of the results. Three of these quasars
(SDSS~J0844$+$0503, SDSS~J1001$+$5454, and SDSS~J1407$+$6454) were
observed as part of our \chandra\ Cycle 7 observations; the remaining
ones have archival \xray\ coverage. Given that the redshifts of our
quasars are sufficiently high (with the possible exception of
SDSS~J0813$+$2545) so that the definitive C~{\sc iv} BAL transition
lies within the spectral range covered by the SDSS spectra (this
requires $z\gtsim1.5$), we do not expect there to be any unidentified
BAL quasars within our sample. Note that the complementary
high-luminosity $z\gtsim4$ quasars from S06 have been chosen to be
radio-quiet, non-BAL quasars.

Three quasars with archival \xray\ observations described in this
paper, SDSS~J0145$-$0945, SDSS~J0813$+$2545, and APM~08279$+$5255, are
gravitationally lensed. SDSS~J0145$-$0945 and APM~08279$+$5255 have
flux-amplification factors of $\approx 3$ and $\approx 100$,
respectively (e.g., Surdej et~al.\ 1988; Egami et~al.\ 2000; Lehar
et~al.\ 2000; E.~O. Ofek 2006, private communication), and we have
calculated the flux-amplification factor of SDSS~J0813$+$2545 to be
$\approx 6$ (based on the $V$-magnitudes taken from the discovery
paper of Reimers et~al. 2002). After correcting for flux amplification
due to lensing, none of these quasars satisfies our $M_i=-29.28$
cutoff. Therefore, we present the basic \xray\ properties of these
three quasars below but exclude them from most of our statistical
analyses. Since our sample consists of sources at the top of the
quasar luminosity function (resulting in a strong magnification bias),
the expected fraction of lensed quasars is of the order of a few
percent (e.g., Turner, Ostriker, \& Gott 1984; E.~O. Ofek 2007,
private communication); this is consistent with the $\sim10\%$
fraction of lensed quasars that we find for our highly luminous
sample. We do not expect unresolved ($\ltsim1''$) gravitational lenses
to be affecting our results materially, and we constrain extended
\xray\ emission in \S2.2.

\subsection{Chandra Observations and Data Reduction}

Our 21 \chandra\ Cycle~7 targets (see \S2.1.1) were observed using the
Advanced CCD Imaging Spectrometer (ACIS; Garmire et~al.\ 2003) with
the aimpoint on the S3 CCD. The requested ``snapshot'' exposure for
each target was 4~ks. All targets were placed near the aimpoint; with
the exceptions of SDSS~J1350$+$5716, SDSS~J1421$+$4633, and
SDSS~J1521$+$5202, all were strongly detected with \hbox{$\approx
10$--150} counts from \hbox{0.5--8\,keV} (details on SDSS~J1521$+$5202
appear in \S4). Faint mode was used for the event telemetry format,
and all observations were free from strong background flares.

Data analysis was carried out using standard {\sc
ciao\footnote{\chandra\ Interactive Analysis of Observations. See
http://asc.harvard.edu/ciao/.} v3.2} routines, and only events with
ASCA grades 0, 2, 3, 4, and 6 were used. {\sc wavdetect} (Freeman et
al.\ 2002) was used for source detection with wavelet scales of 1,
$\sqrt{2}$, 2, $2\sqrt{2}$, and 4 pixels. We adopted a {\sc wavdetect}
false-positive probability threshold of $10^{-4}$. The probability of
spurious detections is very low, given our a priori knowledge of the
locations of our sources and the subarcsecond on-axis resolution of
\chandra. All Cycle~7 targets were detected by \chandra.

We report in Table~3 the counts detected in the ultrasoft band
(\hbox{0.3--0.5\,keV}), the soft band (\hbox{0.5--2\,keV}), the hard
band (\hbox{2--8\,keV}), and the full band
(\hbox{0.5--8\,keV}). Manual aperture photometry with a $3''$-radius
aperture was used to derive the counts. Also in Table~3 we give the
band ratio (the hard-band counts divided by the soft-band counts) and
the effective power-law photon index, $\Gamma$, assuming an \xray\
photon spectrum of the form $N_E \propto E^{-\Gamma}$ across the full
band.  This photon index was calculated from the band ratio using the
\chandra\ {\sc
pimms\footnote{http://cxc.harvard.edu/toolkit/pimms.jsp.} v3.6a} tool;
we used the Cycle~7 instrument response in {\sc pimms} which accounts
for the time-dependent quantum-efficiency decay of ACIS at low
energies (caused by a thin layer of molecular buildup on the ACIS
filter).

\begin{deluxetable*}{lcccccc}
\tablenum{3} \tabletypesize{\scriptsize} \tablecaption{X-ray Counts,
Band Ratios, and Effective Photon Indices of the Core Sample}
\tablehead{ \colhead{} & \multicolumn{4}{c}{X-ray
Counts\tablenotemark{a}} & \colhead{} & \colhead{} \\ \colhead{Object
(SDSS J)} & \colhead{0.3--0.5~keV} & \colhead{0.5--2.0~keV} &
\colhead{2.0--8.0~keV} & \colhead{0.5--8.0~keV} & \colhead{Band
Ratio\tablenotemark{b}} & \colhead{$\Gamma$\tablenotemark{b}} }
\startdata 012156.04$+$144823.9 & 3.0$^{+2.9}_{-1.6}$ &
38.8$^{+7.3}_{-6.2}$ & 6.9$^{+3.8}_{-2.6}$ & 45.8$^{+7.8}_{-6.7}$ &
0.18$^{+0.10}_{-0.07}$ & 2.3$^{+0.5}_{-0.4}$ \\ 014516.59$-$094517.3 &
57.9$^{+8.7}_{-7.6}$ & 566.8$^{+24.8}_{-23.8}$ &
121.3$^{+12.1}_{-11.0}$ & 688.1$^{+27.3}_{-26.2}$ &
0.21$^{+0.02}_{-0.02}$ & 2.1$^{+0.1}_{-0.1}$ \\ 020950.71$-$000506.4 &
$<$ 6.4 & 12.0$^{+4.6}_{-3.4}$ & 5.8$^{+3.6}_{-2.4}$ &
17.8$^{+5.3}_{-4.2}$ & 0.49$^{+0.36}_{-0.24}$ & 1.3$^{+0.6}_{-0.5}$ \\
073502.31$+$265911.4 & $<$ 6.4 & 33.8$^{+6.9}_{-5.8}$ &
6.9$^{+3.8}_{-2.6}$ & 40.7$^{+7.4}_{-6.4}$ & 0.20$^{+0.12}_{-0.08}$ &
2.2$^{+0.5}_{-0.4}$ \\ 075054.64$+$425219.2 & 3.9$^{+3.2}_{-1.9}$ &
45.9$^{+7.8}_{-6.8}$ & 10.9$^{+4.4}_{-3.3}$ & 56.8$^{+8.6}_{-7.5}$ &
0.24$^{+0.10}_{-0.08}$ & 2.0$^{+0.4}_{-0.3}$ \\ 080342.04$+$302254.6 &
$<$ 6.4 & 57.7$^{+8.6}_{-7.6}$ & 13.5$^{+4.8}_{-3.6}$ &
71.2$^{+9.5}_{-8.4}$ & 0.23$^{+0.09}_{-0.07}$ & 2.0$^{+0.3}_{-0.3}$ \\
081331.28$+$254503.0 & 66.0$^{+9.2}_{-8.1}$ & 434.1$^{+21.9}_{-20.8}$
& 159.9$^{+13.7}_{-12.6}$ & 594.0$^{+25.4}_{-24.4}$ &
0.37$^{+0.04}_{-0.03}$ & 1.6$^{+0.1}_{-0.1}$ \\ 084401.95$+$050357.9 &
$<$ 3.0 & 16.0$^{+5.1}_{-4.0}$ & 7.8$^{+3.9}_{-2.8}$ &
23.8$^{+6.0}_{-4.9}$ & 0.49$^{+0.29}_{-0.21}$ & 1.4$^{+0.5}_{-0.4}$ \\
090033.49$+$421546.8 & 8.0$^{+4.0}_{-2.8}$ & 82.0$^{+10.1}_{-9.0}$ &
26.8$^{+6.2}_{-5.2}$ & 108.8$^{+11.5}_{-10.4}$ &
0.33$^{+0.09}_{-0.07}$ & 1.7$^{+0.2}_{-0.2}$ \\ 094202.04$+$042244.5 &
6.0$^{+3.6}_{-2.4}$ & 34.9$^{+7.0}_{-5.9}$ & 11.7$^{+4.5}_{-3.4}$ &
46.6$^{+7.9}_{-6.8}$ & 0.34$^{+0.15}_{-0.11}$ & 1.7$^{+0.4}_{-0.3}$ \\
095014.05$+$580136.5 & $<$ 3.0 & 20.8$^{+5.6}_{-4.5}$ &
5.7$^{+3.5}_{-2.3}$ & 26.6$^{+6.2}_{-5.1}$ & 0.28$^{+0.19}_{-0.13}$ &
1.8$^{+0.6}_{-0.5}$ \\ 100129.64$+$545438.0 & 5.0$^{+3.4}_{-2.2}$ &
55.9$^{+8.5}_{-7.5}$ & 10.7$^{+4.4}_{-3.2}$ & 66.6$^{+9.2}_{-8.1}$ &
0.19$^{+0.08}_{-0.06}$ & 2.1$^{+0.3}_{-0.3}$ \\ 101447.18$+$430030.1 &
3.0$^{+2.9}_{-1.7}$ & 26.0$^{+6.2}_{-5.1}$ & 6.8$^{+3.7}_{-2.6}$ &
32.8$^{+6.8}_{-5.7}$ & 0.26$^{+0.16}_{-0.11}$ & 1.9$^{+0.5}_{-0.4}$ \\
110610.73$+$640009.6 & 17.0$^{+5.2}_{-4.1}$ & 99.9$^{+11.0}_{-10.0}$ &
23.8$^{+6.0}_{-4.9}$ & 123.7$^{+12.2}_{-11.1}$ &
0.24$^{+0.07}_{-0.05}$ & 1.9$^{+0.2}_{-0.2}$ \\
111038.64$+$483115.7\tablenotemark{c} & \nodata & \nodata & \nodata &
\nodata & \nodata & 2.0$^{+0.1}_{-0.1}$ \\
121930.78$+$494052.3\tablenotemark{d} & \nodata & 15.5$^{+5.0}_{-3.9}$
& \nodata & \nodata & \nodata & \nodata \\ 123549.47$+$591027.0 &
6.0$^{+3.6}_{-2.4}$ & 31.0$^{+6.6}_{-5.6}$ & 6.7$^{+3.7}_{-2.5}$ &
37.7$^{+7.2}_{-6.1}$ & 0.22$^{+0.13}_{-0.09}$ & 2.0$^{+0.5}_{-0.4}$ \\
123641.46$+$655442.0 & $<$4.8 & 19.2$^{+5.5}_{-4.4}$ & $<$6.4 &
20.7$^{+5.6}_{-4.5}$ & $<$0.33 & $>$ 1.7 \\ 135044.67$+$571642.8 &
$<$4.8 & 1.9$^{+2.6}_{-1.3}$ & $<$3.0 & 1.7$^{+2.6}_{-1.2}$ & $<$1.56
& $>$2.3 \\ 140747.22$+$645419.9 & 6.0$^{+3.6}_{-2.4}$ &
33.8$^{+6.9}_{-5.8}$ & 12.8$^{+4.7}_{-3.5}$ & 46.7$^{+7.9}_{-6.8}$ &
0.38$^{+.16}_{-0.12}$ & 1.5$^{+0.3}_{-0.3}$ \\ 142123.98$+$463317.8 &
$<$3.0 & 3.9$^{+3.2}_{-1.9}$ & $<$3.0 & 3.7$^{+3.1}_{-1.9}$ & $<$0.77
& $>$0.9 \\ 142656.18$+$602550.9\tablenotemark{d} & \nodata &
8.9$^{+4.1}_{-2.9}$ & \nodata & \nodata & \nodata & \nodata \\
143835.95$+$431459.2 & $<$6.4 & 7.8$^{+3.9}_{-2.8}$ & $<$6.4 &
9.6$^{+4.2}_{-3.1}$ & $<$0.82 & $>$0.9 \\
144542.76$+$490248.9\tablenotemark{d} & \nodata & 28.1$^{+6.4}_{-5.3}$
& \nodata & \nodata & \nodata & \nodata \\ 152156.48$+$520238.4 & $<$
3.0 & $<$ 4.8 & 1.8$^{+2.6}_{-1.3}$ & 2.7$^{+2.9}_{-1.6}$ & $>$ 0.38 &
$<$ 1.5 \\ 152553.89$+$513649.1\tablenotemark{c} & \nodata & \nodata &
\nodata & \nodata & \nodata & 1.8$^{+0.1}_{-0.1}$ \\
161434.67$+$470420.0 & 5.0$^{+3.4}_{-2.2}$ & 126.0$^{+12.3}_{-11.2}$ &
48.8$^{+8.0}_{-7.0}$ & 177.8$^{+14.4}_{-13.3}$ &
0.39$^{+0.07}_{-0.07}$ & 1.5$^{+0.2}_{-0.2}$ \\ 162116.92$-$004250.8 &
$<$ 4.8 & 19.9$^{+5.5}_{-4.4}$ & 6.9$^{+3.8}_{-2.6}$ &
26.8$^{+6.2}_{-5.2}$ & 0.35$^{+0.21}_{-0.15}$ & 1.7$^{+0.5}_{-0.4}$ \\
170100.62$+$641209.0 & 7.9$^{+3.9}_{-2.8}$ & 273.2$^{+17.6}_{-16.5}$ &
74.4$^{+9.7}_{-8.6}$ & 347.6$^{+19.7}_{-18.6}$ &
0.27$^{+0.04}_{-0.04}$ & 1.9$^{+0.1}_{-0.1}$ \\ 173352.22$+$540030.5 &
4.0$^{+3.2}_{-1.9}$ & 32.8$^{+6.8}_{-5.7}$ & 6.9$^{+3.8}_{-2.6}$ &
39.7$^{+7.4}_{-6.3}$ & 0.21$^{+0.12}_{-0.09}$ & 2.1$^{+0.5}_{-0.4}$ \\
212329.46$-$005052.9 & $<$ 4.8 & 21.8$^{+5.7}_{-4.6}$ & $<$ 4.8 &
23.7$^{+5.9}_{-4.8}$ & $<$ 0.22 & $>$ 2.1 \\ 231324.45$+$003444.5 &
$<$3.0 & $<$4.8 & 1.9$^{+2.6}_{-1.3}$ & 2.9$^{+2.9}_{-1.5}$ & $>$0.40
& $<$1.5 \\ \cline{1-7} APM 08279$+$5255 & 66.8$^{+9.2}_{-8.2}$ &
3967.2$^{+64.0}_{-63.0}$ & 1617.4$^{+41.2}_{-40.2}$ &
5584.6$^{+75.6}_{-74.7}$ & 0.41$^{+0.01}_{-0.01}$ &
1.3$^{+0.02}_{-0.03}$ \\ HS 1603$+$3820 & 8.7$^{+4.1}_{-2.9}$ &
93.6$^{+10.7}_{-9.7}$ & 23.2$^{+5.9}_{-4.8}$ & 116.7$^{+11.8}_{-10.8}$
& 0.25$^{+0.07}_{-0.06}$ & 1.7$^{+0.2}_{-0.2}$ \\ \enddata
\tablenotetext{a}{Using Poisson statistics, the errors on the \xray\
counts were calculated according to Tables 1 and 2 of Gehrels (1986)
and correspond to the 1~$\sigma$ level. Upper limits on \xray\ counts
were computed according to Kraft et~al. (1991) and are at the 95\%
confidence level. Upper limits of 3.0, 4.8, and 6.4 correspond to
finding 0, 1, and 2 \xray\ counts within an extraction region of $1''$
centered on the optical position of the quasar, considering the
background negligible within that region.}  \tablenotetext{b}{We
define the band ratio as the hard-band counts divided by the soft-band
counts. Errors for the band ratio and power-law photon index were
calculated at the 1~$\sigma$ level following the ``numerical method''
decribed in \S1.7.3 of Lyons (1991). When the number of counts is
small, this method avoids the failure of the standard
approximate-variance formula. The photon indices have been obtained by
applying the correction required to account for the quantum-efficiency
decay of ACIS at low energies. Note that because the sources in this
sample have been observed in different \chandra\ cycles, the quoted
band ratios cannot be directly compared with each other, due to the
different rest-frame energy bands as well as the time dependence of
the ACIS quantum-efficiency.}  \tablenotetext{c}{The \xray\ properties
(i.e., $\Gamma$ and the soft-band flux) of SDSSJ~1110$+$4831 and
SDSS~J1525$+$5136 have been calculated from \xmm\ observations, which
observes in a different energy band than \chandra. Therefore we do not
have values to quote for the \xray\ counts.}  \tablenotetext{d}{The
\xray\ properties of these objects were determined from soft-band
count measurements taken from archival \rosat\ data.  Therefore we do
not quote values for the \xray\ counts in the other bands.}
\end{deluxetable*}

We examined the data for the presence of extended \xray\ emission
(e.g., due to gravitational lensing or jets) by comparing the radial
profiles of our sources with their expected, normalized
point-spread-functions (PSFs). None of the Cycle~7 targets showed
extended \xray\ emission (but see \S2.1.3). We also did not find
extended \xray\ emission for any of the non-gravitationally lensed
\chandra\ archival quasars.

We checked for an excess of \xray\ companions near our quasars by
searching projected circular regions of $\approx300$ kpc in radius
centered on each source. We found the number of companions within
those fields to be consistent with expectations from the cumulative
number counts from \xray\ surveys (e.g., Bauer et~al.\ 2004).

Rapid variability within the observations of our sources was searched
for by applying Kolmogorov-Smirnov tests to the photon arrival times
of quasars with \gtsim10 counts. No variability was detected, which is
not unexpected given the short exposure times of our observations
(\ltsim30~min in the rest frame).

\subsection{Archival X-ray Data}

Archival data exist and were utilized for 11 of the sources in our
core sample; in Table~1 we give references to past \xray\ studies when
available. The six quasars SDSS~J0145$-$0945 (PI: G.~P.~Garmire),
SDSS~J0813$+$2545 (PI: R.~S.~Priddey), SDSS J1614$+$4704 (PI:
G.~Fossati), SDSS~J1621$-$0042 (PI: J.~Bechtold), SDSS~J1701$+$6412
(PI: L.~P.~Van~Speybroeck), and SDSS~J2313$+$0034 (PI: S.~F.~Anderson)
were observed previously with \chandra. With the exception of
SDSS~J2313$+$0034, all of these sources have $>10$ counts and, aside
from SDSS~J1614$+$4704 and SDSS~J1701$+$6412, have been observed
on-axis. For SDSS~J1614$+$4704 and SDSS~J1701$+$6412, we used
apertures with radii of 30$''$ and 8$''$, respectively, to account for
PSF broadening at large off-axis angles. We averaged the exposure maps
over the same respective apertures when calculating the count
rates. For SDSS~J0145$-$0945, which is a gravitationally lensed
quasar, we used an aperture with a radius of 4$''$ in order to enclose
the counts from both images. Although SDSS~J0813$+$2545 is also
gravitationally lensed, the angular separation between the images is
small enough for a $3''$-radius aperture to suffice. Data reduction
for these sources was carried out in a similar manner to that
described in \S2.2, with the proper correction applied to each source
for the time-dependent quantum-efficiency decay of ACIS at low
energies.

Archival \rosat\ data were used for three of the quasars:
SDSS~J1219$+$4940 (PI: J.~Liebert), SDSS~J1426$+$6025 (PI:
D.~Reimers), and SDSS~J1445$+$4902 (PI: U.~Herbstmeier). These objects
were detected by the \rosat\ PSPC instrument, with 15.5, 8.9, and 28.1
counts, respectively, in the 0.5$-$2.0 keV band. SDSS~J1219$+$4940 and
SDSS~J1445$+$4902 have off-axis angles of 20.6$'$ and 16.1$'$, while
SDSS~J1426$+$6025 was observed on-axis. The counts were derived for
these quasars using aperture sizes of 45$''$ for SDSS~J1426$+$6025,
140$''$ for SDSS~J1219$+$4940, and 100$''$ for SDSS~J1445$+$4902;
background was estimated by placing circular apertures on regions
where no other \xray\ sources were present. Using {\sc pimms} to
extrapolate our measured \hbox{$0.5$--2.0~keV} count rate for
SDSS~J1426$+$6025 to the full \rosat\ energy range ($\approx
0.1$--2.4~keV), we found our value was consistent with that previously
published by Reimers et~al. (1995).

The remaining two archival sources in our sample, SDSS~J1110$+$4831
and SDSS~J1525$+$5136, were previously observed with
\xmm. SDSS~J1110$+$4831 has its \xray\ properties described in detail
in Page et~al.\ (2005). We have re-analyzed the \xmm\ data using
standard \xmm\ Science Analysis System {\sc v6.5.0} tasks and found
results consistent with those presented in Page et~al.\ (2005); the
\xray\ values we quote in Tables 3 and 4 are taken from our
analysis. Similarly, the \xray\ data for SDSS~J1525$+$5136 have
appeared in Page et~al.\ (2005), and this quasar has been studied in
detail in Shemmer et~al.\ (2005b); we use the data from Shemmer
et~al.\ (2005b) in our analysis.

\section{X-ray, Optical, and Radio Properties of the Core Sample}

In Table\,\ref{tab4} (placed at the end of the paper) we list the main
X-ray, optical, and radio properties of our core sample:\smallskip \\
{\sl Column (1)}. --- The SDSS J2000.0 quasar coordinates, accurate to
$\sim$ 0.1$''$.\smallskip \\ {\sl Column (2)}. --- Galactic column
density in units of 10$^{20}$ cm$^{-2}$, calculated using {\sc
colden\footnote{See http://cxc.harvard.edu/toolkit/colden.jsp.}} with
the data from Stark et~al.\ (1992).\smallskip \\ {\sl Column (3)}. ---
The monochromatic {\it AB} magnitude at a rest-frame wavelength of
1450~\AA\ ({\it AB}$_{1450}$ = $-$2.5log $f_{\mbox{\scriptsize
1450~\AA}}$ $-$ 48.6; Oke \& Gunn 1983). The {\it AB}$_{1450}$
magnitudes were calculated from the spectra after applying corrections
due to Galactic extinction and fiber light-loss. The fiber light-loss
correction was calculated as the average difference between the
synthetic $g$, $r$, and $i$ magnitudes (i.e., the integrated flux
across each respective bandpass in the SDSS spectrum) and the
photometric $g$, $r$, and $i$ magnitudes, assuming no flux variation
between the spectroscopic and photometric epochs. The {\it
AB}$_{1450}$ magnitudes for the two quasars that were missed by the
SDSS selection criteria, APM~08279$+$5255 and HS~1603$+$3820, were
calculated from the photometric $i$ magnitudes.
\smallskip \\ {\sl Column (4)}. --- The absolute {\it i}-band
magnitude, taken from the SDSS DR3 quasar catalog; for
APM~08279$+$5255 and HS~1603$+$3820, we calculated this value from the
photometric $i$-magnitude, correcting for Galactic
extinction.\smallskip \\ {\sl Columns (5) and (6)}. --- The flux
density and luminosity at a rest-frame wavelength of 2500~\AA\,
computed from the magnitude in Column 3, assuming a UV-optical
power-law slope of $\alpha=-0.5$ (Vanden Berk et~al. 2001), where
$F_\nu\propto\nu^{\alpha}$.\smallskip \\ {\sl Columns (7) and
(8)}. --- The count rate in the observed-frame \hbox{0.5--2.0\,keV}
band and the corresponding flux, corrected for Galactic absorption and
the quantum-efficiency decay of \chandra\ ACIS at low energy. The
fluxes have been calculated using {\sc pimms}, assuming a power-law
model with $\Gamma$ = 2.0, which is a typical photon index for
luminous AGNs (e.g., Reeves \& Turner 2000; Page et~al.\ 2005;
Piconcelli et~al.\ 2005; Shemmer et~al.\ 2005a; Vignali et~al.\ 2005;
see \S5.1 for direct justification).\smallskip \\ {\sl Columns (9) and
(10)}. --- The flux density and luminosity at a rest-frame energy of
2\,keV, computed assuming $\Gamma$ = 2.0 and corrected for the
quantum-efficiency decay of \chandra\ ACIS at low energy.\smallskip \\
{\sl Column (11)}. --- The luminosity in the rest-frame
\hbox{2$-$10\,keV} band.\smallskip \\ {\sl Column (12)}. --- The
\xray-to-optical power-law slope, \aox, defined as:

%%
%% FIGURE 2
%%
\begin{figure*}
\plotone{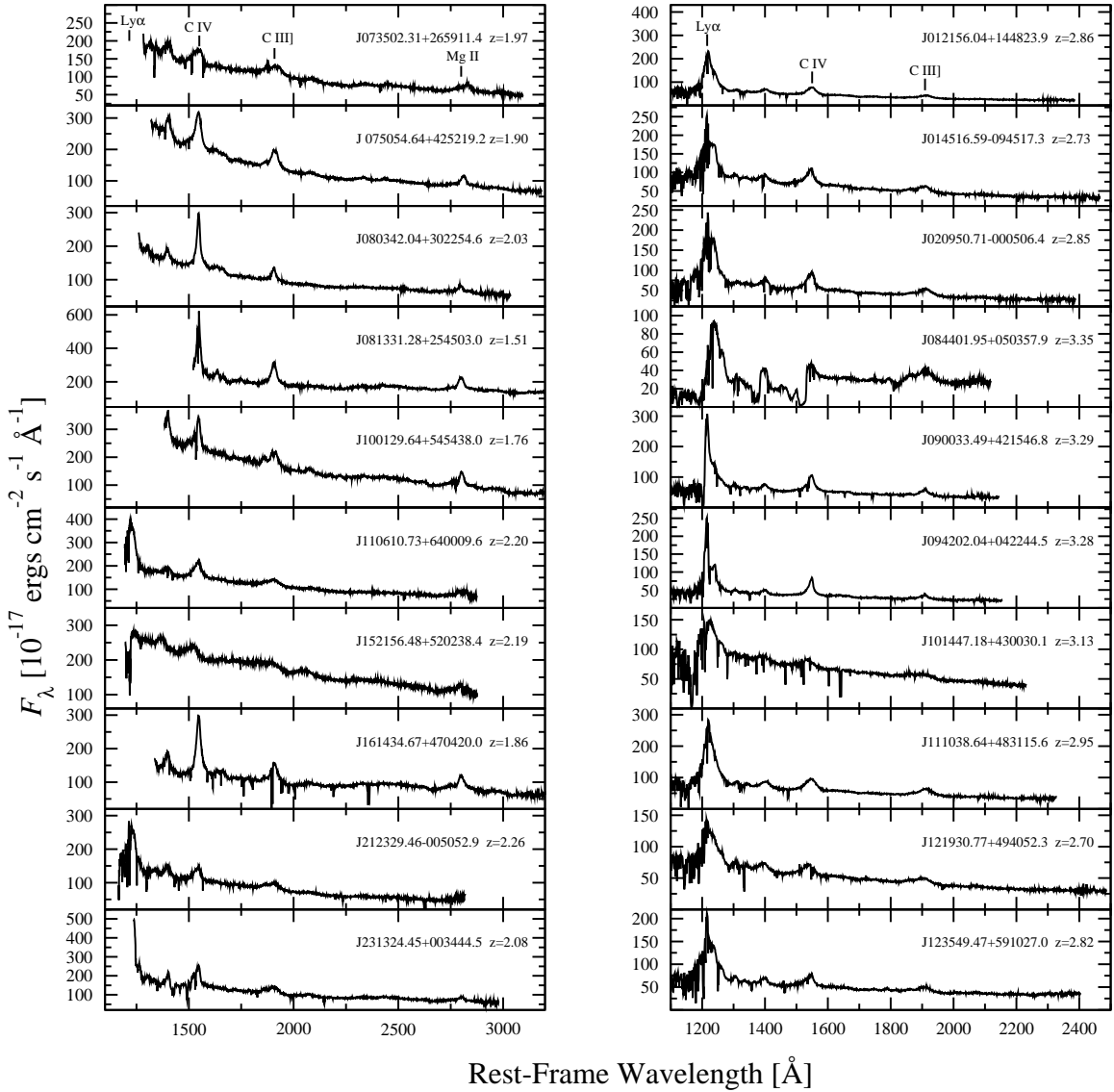}
\figurenum{2}
%\vskip -6.0cm
\caption{\label{sdss_spectra} SDSS spectra for our sample of 32 SDSS
 quasars. Prominent emission lines are marked in the top panel of each
 column. The spectral resolution is $\approx 2000$. The four columns
 have been sorted into redshift bins in the following way: the first
 contains sources from \hbox{$z\sim1.5$--2.5}, the second and third
 columns from \hbox{$z\sim2.5$--3.5}, and the fourth column from
 \hbox{$z\sim3.5$--4.5}. Within each column the spectra are sorted by
 right ascension. SDSS~J0844$+$0503, SDSS~J1525$+$5136 and
 SDSS~J2313$+$0034 are BAL quasars. Note the C~{\sc iv} absorption for
 SDSS~J0844$+$0503, which is also mildly radio-loud, and the
 interesting spectrum for SDSS~J1521$+$5202, which is shown in greater
 detail in Figure~\ref{1521_spectrum}.  }
\label{images1}
\end{figure*}

\begin{figure*}
\figurenum{2}
\plotone{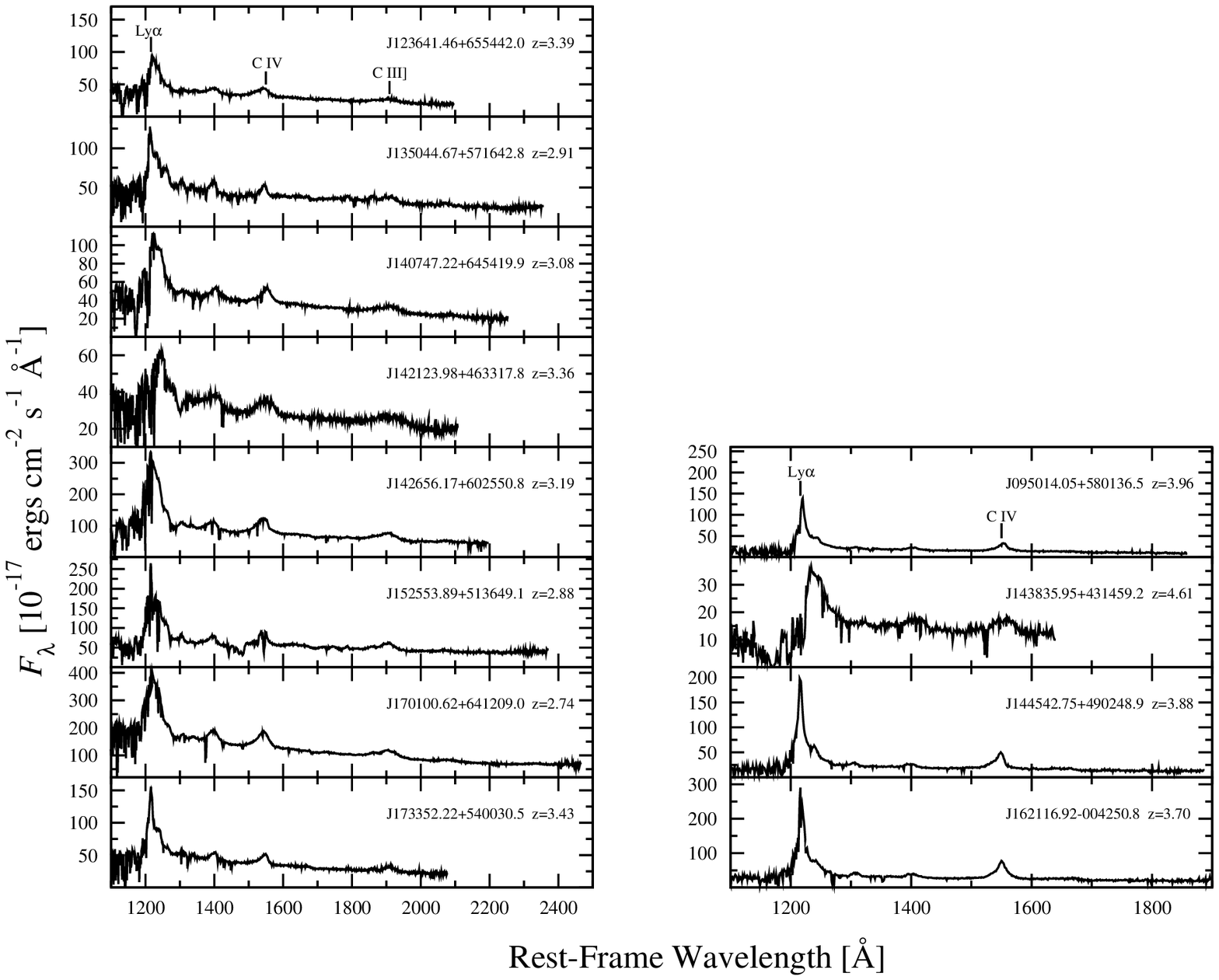}
\caption{{\it Continued}}
\end{figure*}

\begin{equation}
\alpha_{\rm ox} \equiv \frac{{\rm log}(f_{2~\rm
keV}/f_{\mbox{\scriptsize 2500~\AA}})} {{\rm log}(\nu_{2~\rm
keV}/\nu_{\mbox{\scriptsize 2500~\AA}})} = 0.3838~{\rm log}(f_{2~\rm
keV}/f_{\mbox{\scriptsize 2500~\AA}})\ \smallskip \\
\end{equation} where $f_{2~\rm keV}$ and $f_{\mbox{\scriptsize 2500~\AA}}$
are the monochromatic flux densities at
rest-frame 2\,keV and 2500~\AA, respectively.\smallskip \\ {\sl Column
  (13)}. --- The difference between the measured \aox\ (from
Column\,12) and the predicted \aox\ (quoted as \daox), given the UV
luminosity from Column\,6, based on the established \aox--\luv\
relation [given as equation (2) of S06]. The statistical significance of
this difference is also given in units of $\sigma$, where
$\sigma=0.146$ for \hbox{$31<${\rm log}(\luv)$<32$}, and
$\sigma=0.131$ for \hbox{$32<${\rm log}(\luv)$<33$} (see Table~5 of
S06). \smallskip \\ {\sl Column (14)}. --- The radio-loudness
parameter $R$ (see \S2.1.3).  The $f_{\mbox{\scriptsize4400~\AA}}$
values were calculated by extrapolating from the magnitudes in column
3, assuming a UV-optical power-law slope of $\alpha=-0.5$. The $f_{5
  \rm{GHz}}$ values were calculated using data from the FIRST (Faint
Images of the Radio Sky at Twenty cm; Becker et~al.\ 1995) and NVSS
(NRAO VLA Sky Survey; Condon et~al.\ 1998) catalogs. The flux density
at a rest-frame frequency of 5~GHz was calculated from the flux
density at an observed-frame frequency of 1.4~GHz assuming a radio
power-law slope of $\alpha=-0.8$. Seven of our sources have FIRST
radio detections; two of them are RLQs. Upper limits were placed on 20
of our sources at the 3~$\sigma$ level, given our a priori knowledge
of the positions of all of our sources. Upper limits on the five
sources not covered by the FIRST survey were placed using the upper
limit on NVSS detection ($\approx 2.5$~mJy).

For the luminosity values quoted in Columns 4, 6, 10, and 11, no
lensing corrections have been made (i.e., the fluxes have not been
de-amplified when determining these values).

\section{Optical Spectra and Notes on Individual Objects}

In Figure~\ref{sdss_spectra} we present the optical spectra of the 32
SDSS quasars in our sample. Below we comment on sources with
particularly interesting optical and/or \xray\ properties.

\begin{figure}
\epsscale{0.8}
\plotone{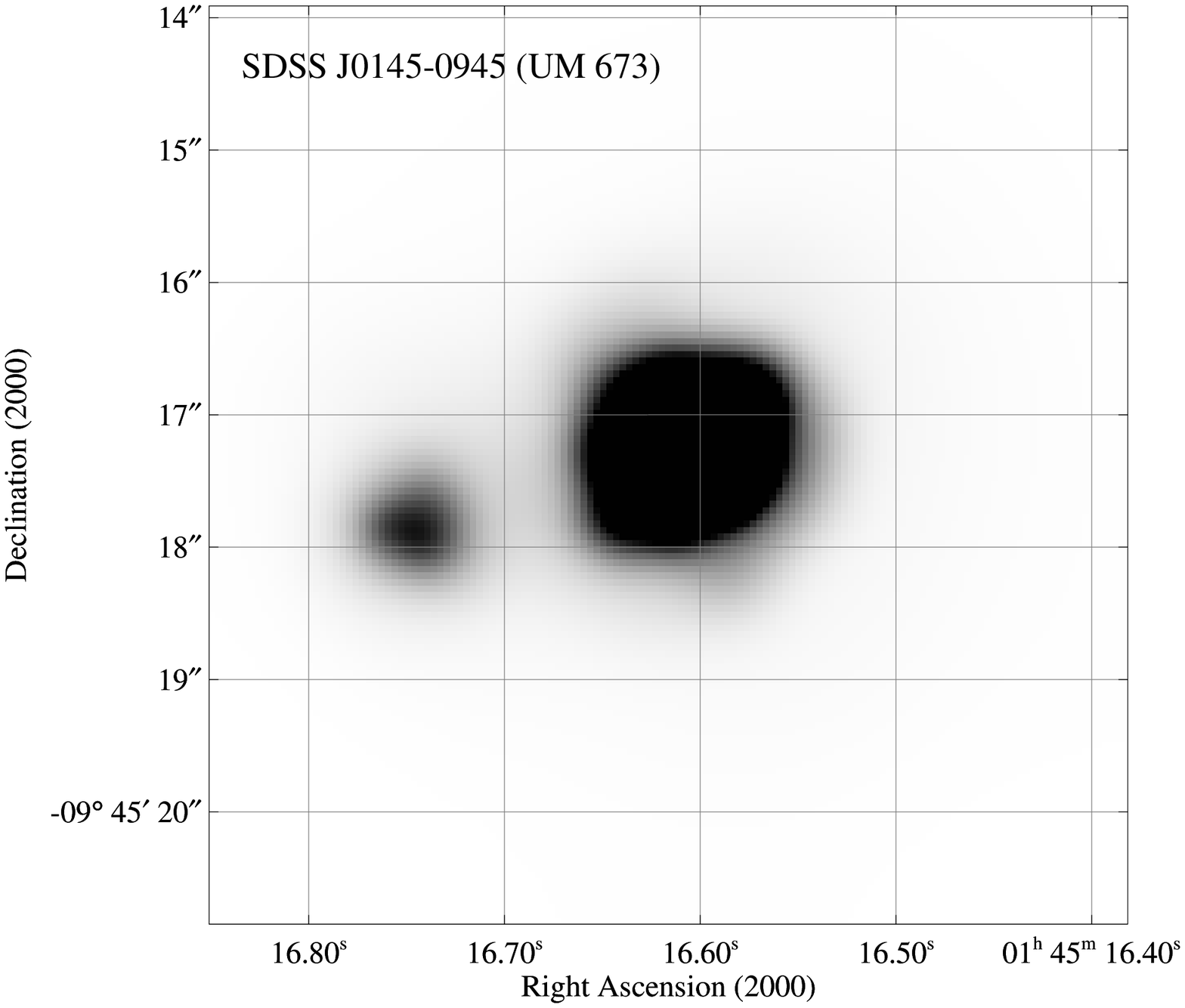} 
\figurenum{3}
\caption{\label{xray_image} \chandra\ \hbox{0.5--8\,keV} image of the
 gravitationally lensed quasar SDSS~J0145$-$0945. The image spans
 \hbox{$\approx 7'' \times 7 ''$} on the sky; North is up, and East is
 to the left. The image has been adaptively smoothed at the 2~$\sigma$
 level. The two quasar images are separated by $\Delta\theta\approx
 2''$, and the optical and \xray\ flux ratios between them are
 $\approx 7$ and $\approx 12$, respectively (note that the optical and
 \xray\ flux measurements were taken at different epochs). The
 positions of the two \xray\ images agree with those of the optical
 images (based upon SDSS astrometry) to within the expected \chandra\
 positional error.  }
\end{figure}

{\sl SDSS~J014516.59$-$094517.3} ($z=2.73$).---This object was first
reported as a gravitationally-lensed system by Surdej et~al.\ (1987,
1988), and it consists of two images separated by $\Delta\theta
\approx 2''$. The optical flux ratio between the two images was found
to be $\approx 7$ (Kassiola \& Kovner 1992). Using the \chandra\
observations taken in 2004 we have calculated an \xray\ flux ratio of
$12.4^{+2.0}_{-1.7}$ in the full band (see Figure~\ref{xray_image});
this discrepancy between flux ratios is not wholly unexpected, given
(1) the amplitude of flux variations in the \xray\ band is typically
greater than that in the optical, (2) the likelihood of intrinsic
variability of the quasar between the different epochs, and (3)
possible microlensing. The optical positions measured from SDSS
astrometry agree with the positions of the \xray\ centroids of both
components to within the expected astrometric accuracy of \chandra\
($\approx0.5''$). We found \hbox{\aox~$=-1.51$} for this quasar, which
differs from the predicted value from S06 (using a lensing-corrected
luminosity) by 1.3~$\sigma$.

{\sl SDSS~J081331.28$+$254503.0} ($z=1.51$).---This object is a
gravitationally-lensed system consisting of four images discovered by
Reimers et~al.\ (2002). The angular separation between the two
brightest images is $\Delta\theta=0.25''$, too small to be resolved
by the angular resolution of \chandra\ ($\approx0.5''$). We measure
\hbox{\aox~$-1.58$} for this quasar, which differs from the predicted
value of S06 (using a lensing-corrected luminosity) by 0.4~$\sigma$.

{\sl SDSS~J084401.95$+$050357.9} ($z=3.35$).---This is a BAL RLQ, with
a relatively mild radio-loudness parameter of $R=18.9$ and a
rest-frame equivalent width (EW) of $\approx \rm{30~\AA}$ for the
C~{\sc iv} absorption trough. While the additional jet-linked \xray\
emission generally increases the \xray-to-optical flux ratio, the
intrinsic absorption present in BAL quasars tends to reduce that
quantity; it is possibly because of these competing effects that we
measure \hbox{\aox~$=-1.72$} for this source, which differs from the
predicted value for a non-BAL RQQ with log(\luv)~$=31.90$ at a level
of only 0.07~$\sigma$.

{\sl SDSS~J135044.67$+$571642.8} ($z=2.91$).---This quasar is somewhat
\xray\ weak, with \aox~$=-2.14$; this value differs from the predicted
value by 3.02~$\sigma$. The SDSS spectrum of this object appears in
Figure~\ref{sdss_spectra}, and it shows no obvious UV~absorption.

\begin{figure}
\epsscale{0.8}
\plotone{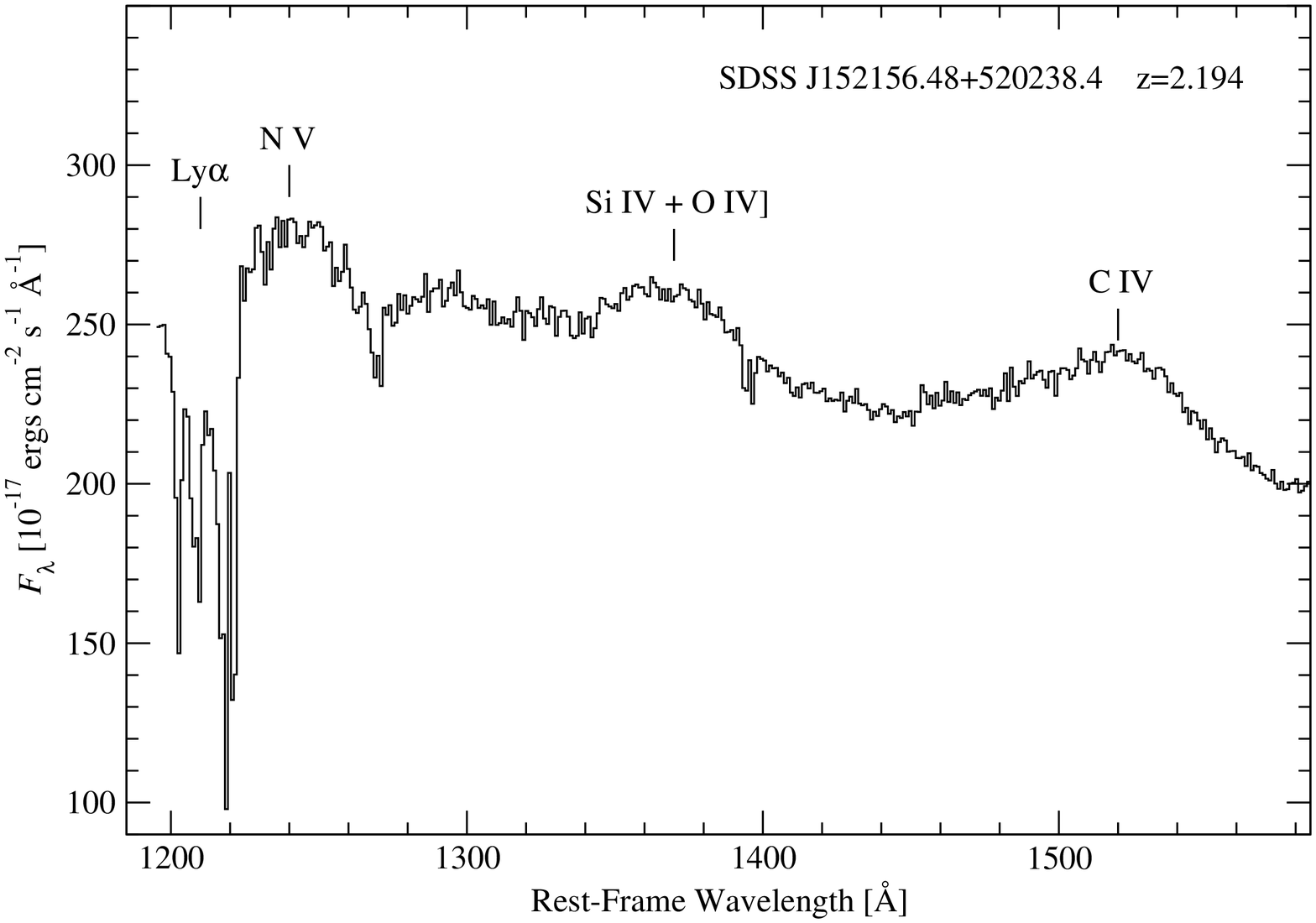} 
\figurenum{4}
\caption{\label{1521_spectrum}
 The SDSS spectrum of SDSS~J1521$+$5202 (see \S4). This object shows strong
 Ly$\alpha$ absorption, which suggests that absorbing material along the
 line-of-sight is responsible for its anomalous \xray\ weakness (\aox$=-2.44$).
 The spectral resolution is $\approx 2000$.
}
\end{figure}

{\sl SDSS~J152156.48$+$520238.4} ($z=2.19$\footnote{Although the SDSS
quotes a redshift of 2.21, upon examination of the spectrum we have
measured a redshift of 2.19 based upon the Mg~{\sc ii} $\lambda2798$
line; it is this latter value which we adopt throughout this
paper.}).---This exceptionally luminous quasar is the third most
optically luminous object in S05. Manual aperture photometry for this
quasar measured only 3 counts: 2 in the hard band and 1 in the soft
band. \xray\ fluxes and other properties were calculated from the
full-band count rate using {\sc pimms} since there were not enough
counts in the soft band for a detection. This quasar is anomalously
\xray\ weak, with log($\nu L_\nu$)$_{\rm 2~keV}=43.74$ and a steep
\aox~$=-2.44^{+0.12}_{-0.15}$, which is inconsistent with the
predicted value at a level of $\approx 5$~$\sigma$. The SDSS spectrum
of this quasar appears in Figure~\ref{1521_spectrum}. The Ly$\alpha$
line is completely absorbed by several narrow absorption-line (NAL)
systems, and the high-ionization emission lines are blueshifted
relative to the quasar's redshift, even when allowing for the revised
redshift in Footnote~8. The strong observed UV absorption and the hint
of a hard \xray\ spectral shape suggest that absorbing material along
the line of sight is likely responsible for the \xray\ weakness of
this quasar (e.g., Brandt, Laor, \& Wills 2000; Gallagher et~al.\
2001). Therefore, we will exclude this quasar from the statistical
analyses below, since our main interest there is in the intrinsic
\xray\ emission properties of quasars.

{\sl SDSS~J170100.62$+$641209.0} ($z=2.74$).---This quasar is the most
optically luminous in the DR3 catalog and has an \aox~$-1.91$, which
differs from the predicted value by 0.7~$\sigma$. After binning the
data from the \xray\ spectrum into 10 full-band counts per bin, we
fitted the spectrum below 2~keV to a power-law model with Galactic
absorption and extrapolated the fit to higher energies; we found two
consecutive bins that fall $\gtsim 3~\sigma$ below the model near
$\sim 3$~keV in the observed frame ($\sim 11$~keV in the rest frame),
signs of a possible absorption feature. A spectrum taken with \xmm\
(PI: F.~Jansen) shows no such feature, although it is not ruled out
within the uncertainties on the data points (the \xmm\ spectrum was
particularly noisy due to background flaring). The binned \chandra\
spectrum of this quasar appears as part of
Figure~\ref{ind_xray_spectra}. Further observations are necessary to
test the reality of this feature.

%%
%% FIGURE 3
%%

\section{Data Analysis and Results}

\subsection{X-ray Spectral Properties}

\subsubsection{Individual-Object Analyses}

We have investigated the \xray\ spectra of seven of the eight sources
with \chandra\ observations that have $> 100$ full-band counts
(APM~08279$+$5255 has already had its complex spectrum studied in
detail in Chartas et~al.\ 2002); below this threshold there are too
few counts for statistically useful results to be derived from
individual spectral fitting. The seven spectra were extracted with the
CIAO routine {\sc psextract} using circular apertures of $3.0''$ in
radius centered on the \xray\ centroid of each source, with the
exceptions of SDSS~J0145$-$0945 ($4.0''$ used to enclose both lensed
images), SDSS~J1614$+$4704 ($30.0''$ used due to PSF broadening at
large off-axis angles), and SDSS~J1701+6412 ($8.0''$ used due to PSF
broadening). Background regions were extracted using annuli of varying
sizes to avoid contamination from other \xray\ sources. An annulus was
not used to extract the background for SDSS~J1701$+$6412 due to the
quasar's location near the edge of the ACIS I3 CCD; instead we chose a
nearby circular region $20''$ in radius that was free from other
\xray\ sources.

We used XSPEC~{\sc v11.3.2} (Arnaud 1996) to fit each spectrum across
the full-band energy range (\hbox{0.5--8.0 keV}) with a power-law
model and a fixed Galactic-absorption component (Dickey \& Lockman
1990); all fits assumed solar abundances (e.g., Anders \& Grevesse
1989) and used the {\sc wabs} absorption model in XSPEC. We used the
$C$-statistic (Cash 1979) when modelling the unbinned data, since this
method is more appropriate when fitting low-count sources than
$\chi^2$ fitting and still remains accurate for higher numbers of
counts (e.g., Nousek \& Shue 1989). All of the errors have been quoted
at the 90\% confidence level considering one parameter to be of
interest ($\Delta C=2.71$; Avni 1976; Cash 1979). Although when using
the $C$-statistic there is no value analogous to $P(\chi^2|\nu)$ with
which to perform model testing, we assessed whether each model fits
the data acceptably by searching for any systematic residuals. The
seven objects with fitting, along with their fit parameters and
statistics, appear in Table~5. In Figure~\ref{ind_xray_spectra} we
present their \xray\ spectra, binned at a level of 10 counts per bin
for clearer presentation. Note that in Figure~\ref{ind_xray_spectra}
(unlike in Table~5) we used $\chi^2$ fitting in order to show
residuals in units of $\sigma$. The values of $\Gamma$ calculated from
the band ratios (see Table~3) are consistent with those derived from
the best-fit models. We also added an intrinsic, redshifted,
neutral-absorption component to the model, but it did not
significantly improve any of the fits.
%%
%% FIGURE 4
%%

\begin{deluxetable*}{lcccccc}
\tablecolumns{7} \tablenum{5} \tabletypesize{\scriptsize}
\tablecaption{Individual-Object Spectral-Fitting
Results\tablenotemark{a}} \tablehead{ \colhead{Object} &
\colhead{\xray} & \colhead{} & \colhead{$C$\tablenotemark{b}} &
\colhead{} & \colhead{$N_{\rm H}$\tablenotemark{c}} &
\colhead{$C$\tablenotemark{c}} \\ \colhead{(SDSS~J)} &
\colhead{Counts} & \colhead{$\Gamma$\tablenotemark{b}} &
\colhead{Statistic} & \colhead{$\Gamma$\tablenotemark{c}} &
\colhead{(10$^{22}{\rm cm}^{-2}$)} & \colhead{Statistic} } \startdata
0145$-$0945 & 686 & 2.05$^{+0.12}_{-0.12}$ & 151.7 &
2.08$^{+0.20}_{-0.15}$ & $\le 1.00$ & 151.4 \\ 0813$+$2545 & 591 &
1.65$^{+0.12}_{-0.12}$ & 195.4 & 1.65$^{+0.11}_{-0.11}$ & $\le
0.16$\tablenotemark{d} & 195.2 \\ 0900$+$4215 & 108 &
1.94$^{+0.29}_{-0.28}$ & 74.5 & 1.94$^{+0.30}_{-0.28}$ & $\le 1.56$ &
74.5 \\ 1106$+$6400 & 122 & 2.01$^{+0.28}_{-0.28}$ & 73.7 &
2.01$^{+0.37}_{-0.27}$ & $\le 1.18$ & 73.7 \\ 1614$+$4704 & 181 &
1.74$^{+0.28}_{-0.22}$ & 88.1 & 1.86$^{+0.35}_{-0.32}$ & $\le 1.80$ &
88.0 \\ 1701$+$6412 & 352 & 1.91$^{+0.16}_{-0.16}$ & 153.3 &
1.91$^{+0.22}_{-0.16}$ & $\le 1.47$ & 153.3 \\ HS~1603$+$3820 & 116 &
2.01$^{+0.29}_{-0.28}$ & 63.2 & 2.10$^{+0.51}_{-0.35}$ & $\le 1.29$ &
62.8 \\ \enddata \tablenotetext{a}{All fits are across observed-frame
$0.5$--8.0~keV and include appropriate Galactic absorption.}
\tablenotetext{b}{Without an intrinsic absorption component.}
\tablenotetext{c}{With an intrinsic absorption component. Upper limit
on the intrinsic column density at the source redshift given in Table
1, quoted at the 90\% confidence level.}  \tablenotetext{d}{The
tighter constraint on this intrinsic column density is due to the
relatively low redshift ($z=1.51$) of this object combined with its
relatively large number of counts.}
\end{deluxetable*}

SDSS~J1701$+$6412 appears to show an absorption feature at $\sim
3$~keV in the observed frame (see \S4). We used $\chi^2$ fitting to
investigate further the significance of this feature; the $\approx
350$ full-band counts detected from this object are enough for
$\chi^2$ fitting to be acceptable. When modelled with a power law and
Galactic absorption, $P(\chi^2|\nu)=0.23$ with $\chi^2=34.1$ and
$\nu=29$. Although this is a statistically acceptable fit, the
presence of systematic residuals motivated further
investigation. Performing an $F$-test, we found the addition of
intrinsic absorption to the model did not significantly improve the
fit. However, the addition of an absorption edge at 2.4~keV in the
observed frame (9.0~keV in the rest frame) significantly improved the
fit at a confidence level greater than 99.6\% ($\Delta\chi^2=11.8$ for
2 additional fit parameters). Fe~{\sc xxv} has an ionization energy of
8.8~keV, which is close to the location of the modelled edge. Further
observations are required to assess better the nature of the \xray\
spectral complexity in this remarkably luminous quasar.

%%
%% FIGURE 5
%%
\begin{figure}
%\epsscale{0.8}
\plotone{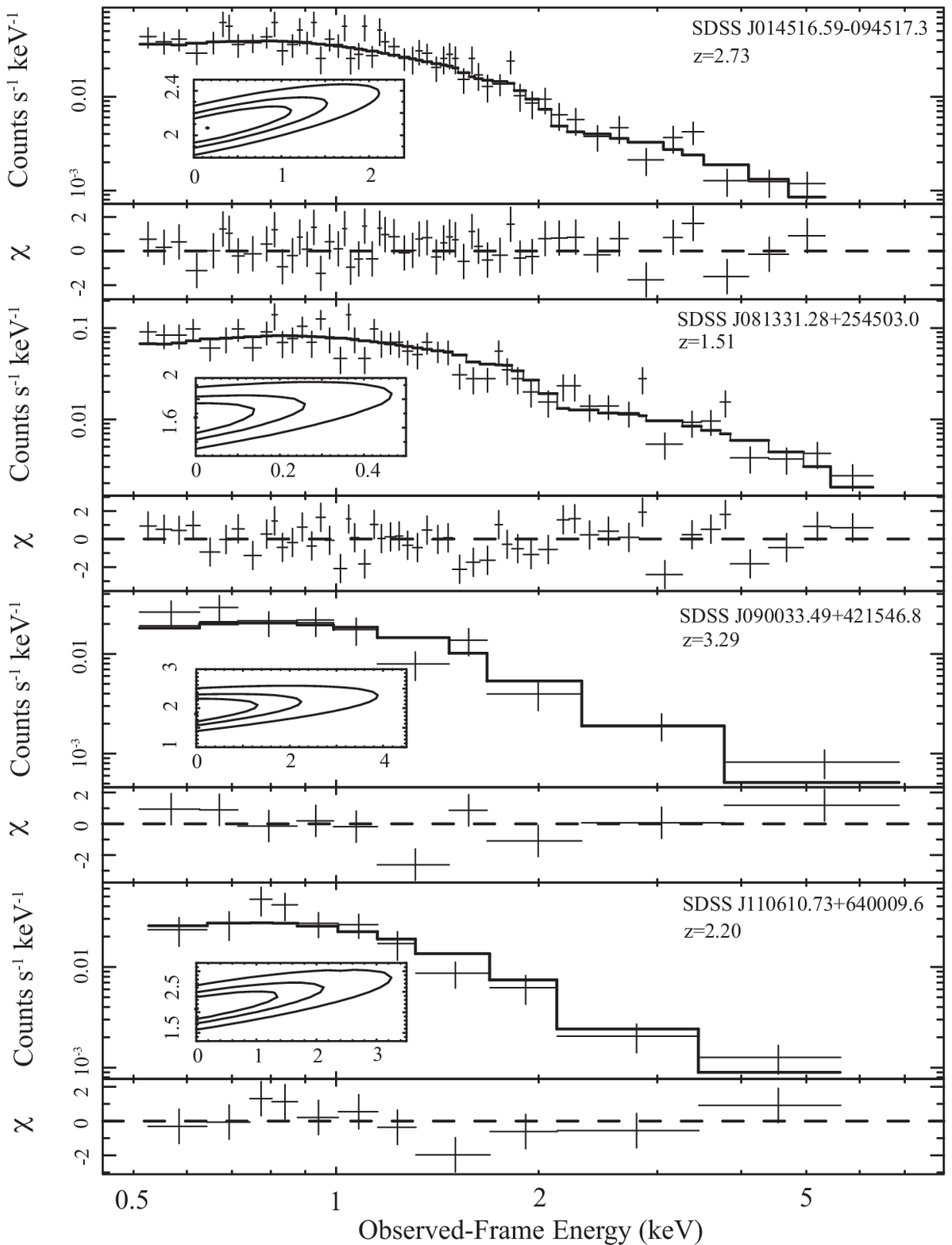} 
\figurenum{5}
\caption{\label{ind_xray_spectra} Individual \chandra\ \xray\ spectra
 and residuals for seven of the quasars from our sample which have
 $>100$ counts. The spectra have been fitted across the full band
 \hbox{($0.5$--8~keV)}. The $\chi$ residuals are in units of $\sigma$,
 and the inset in each panel shows a contour map of $\Gamma$ versus
 intrinsic $N_{\rm H}$ (in units of $10^{22}~{\rm cm}^{-2}$) at
 confidence levels corresponding to 68\%, 90\%, and 99\%.  }
\end{figure}
\begin{figure}
%\epsscale{0.8}
\plotone{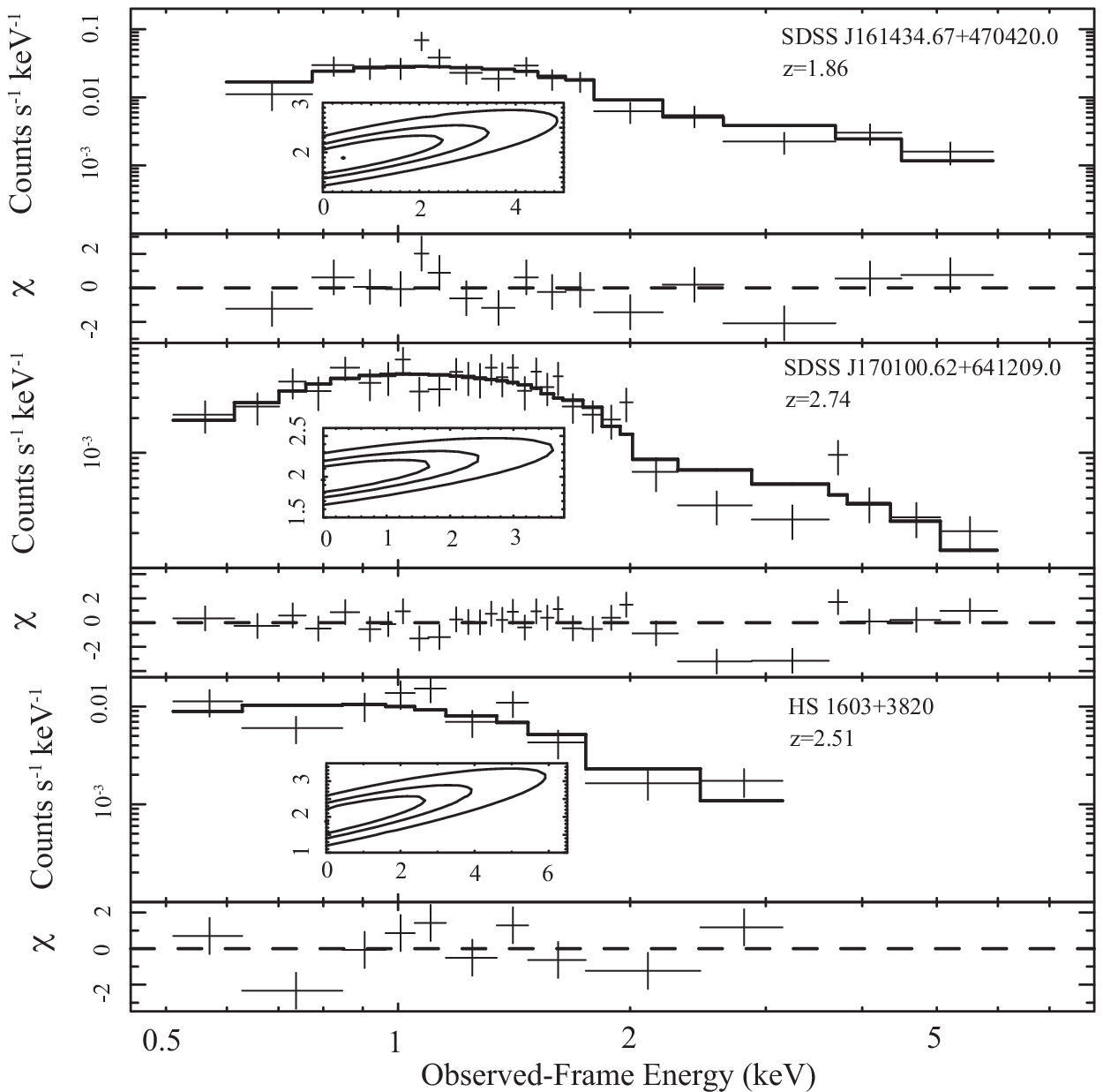}
\figurenum{5}
\caption{{\it Continued}}
\end{figure}

\subsubsection{Joint-Spectral Analyses}

The relatively small numbers of counts for many of the sources in our
sample make it impossible to measure accurately $\Gamma$ and $N_{\rm
H}$ on a source-by-source basis, so joint fitting has been used for
different combinations of sources in our core sample combined with the
complementary $z\gtsim4$ sources from S06 (see \S2.1.2). This
procedure allows measurement of the average \xray\ properties of sets
of objects, including those with too few counts for individual
spectral fitting. Only 42 sources that have \chandra\ observations
have been used in the joint fitting. Shown in Table~6 are the results
of the joint fits across the observed $0.5$--8.0~keV range, all of
which exclude gravitationally-lensed (since their deamplified fluxes
do not meet our luminosity criterion), radio-loud, and BAL
quasars. Also excluded is SDSS~J1521$+$5202 (see \S4), as well as
sources which had less than 3 full-band counts; these include
SDSS~J1350$+$5716 (with 2 counts) and two of the complementary
$z\gtsim4$ sources from S06 (BR~1117$-$1329, with 2 counts, and
PSS~2344$+$0342, which was not detected in the \xray\ band). We
extracted the X-ray spectra of these sources using {\sc psextract} in
a similar manner to that described in \S5.1.1. The sets of unbinned
spectra were fit using XSPEC, first with a power-law model and a
Galactic absorption component, which was kept fixed during the fit,
and then with an added intrinsic neutral-absorption component. All
fits utilized the $C$-statistic.

We checked whether our sample was biased by objects with a high
signal-to-noise ratio (S/N). We split the quasars into two groups,
each having more and less than 100 full-band counts. It can be seen
from Table~6 that the $\Gamma$ found for the quasars with $>100$
counts is consistent with the $\Gamma$ found for quasars with $<100$
counts; we therefore conclude that the high S/N spectra do not bias
the overall sample (note that the total number of counts is
approximately split between the two groups). The addition of an
intrinsic-absorption component did not improve any of the fits; i.e.,
no significant amount of intrinsic absorbing material has been
detected. Upper limits on intrinsic $N_{\rm H}$ appear in Table~6.

\begin{figure}
%\epsscale{0.8}
\plotone{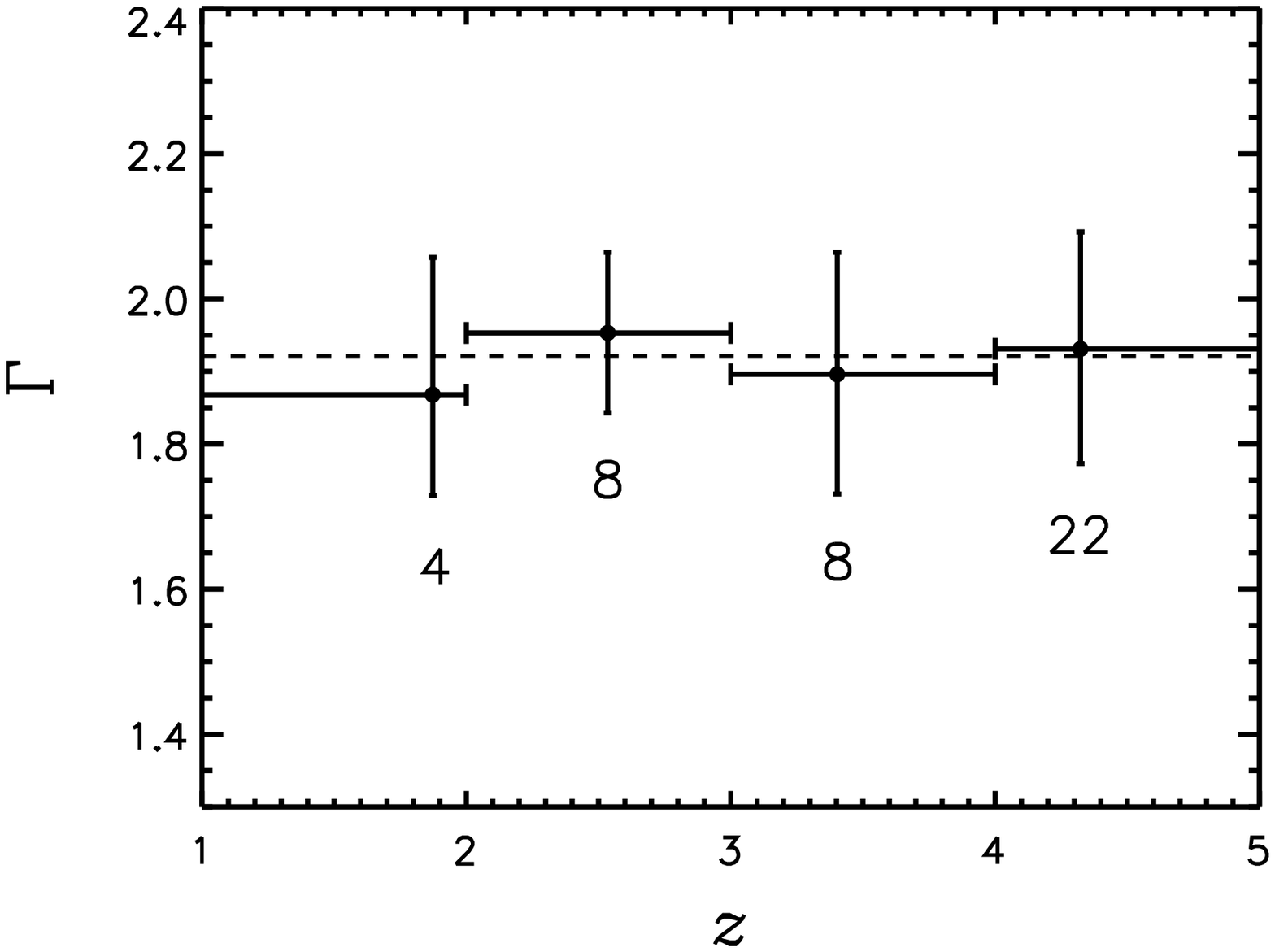} 
\figurenum{6}
\caption{\label{gamma_plot} \xray\ power-law photon index ($\Gamma$)
 vs. redshift, binned into four integer redshift bins ranging from
 $z=1$--5, for our sample and the complementary high-luminosity
 sources at $z\gtsim4$ from S06 (see \S2.1.2). The mean photon indices
 were derived from joint spectral fitting across the full band
 ($0.5$--8~keV). Only sources with \chandra\ observations are
 included, while gravitationally lensed, radio-loud, and BAL quasars
 have been excluded (as well as SDSS~J1521$+$5202). The vertical error
 bars show 90\% confidence bars in each bin, and the number of sources
 in each bin is shown beneath the error bars; three sources with $<3$
 counts (SDSS~1350$+$5716, BR~1117$-$1329, and PSS~2344$+$0342) did
 not meet our requirements for joint fitting and thus were excluded
 (see \S5.1.2). The horizontal error bars show the width of each bin,
 and the data points are marked at the median redshift in each
 bin. The dashed line shows the best constant-model fit, which has a
 value of 1.92.  }
\end{figure}

Previous studies have shown that, in general, $\Gamma$ does not evolve
with redshift for luminous quasars (e.g., Page et~al.\ 2005; Shemmer
et~al.\ 2005a, 2006a; Vignali et~al.\ 2005). To investigate this
matter further, we have performed joint fitting on sets of our quasars
binned into four integer redshift bins ranging from $z=1$--5. Our
sample of the most optically luminous quasars spans the full redshift
range where such objects are known to exist in the Universe,
\hbox{$z\approx 1.5$--4.5}, and it is constrained to a relatively
narrow luminosity range [having a mean log(\luv)~$=32.2$; no mean
\luv\ in any integer redshift bin differs from this global mean by
more than $2~\sigma$, with three of the four redshift bins differing
by less than $1~\sigma$]. These two properties enable our sample to
explore a different region of the luminosity-redshift plane than
previous studies, and they minimize possible confusion between
redshift-dependent and luminosity-dependent effects. For the joint
fitting we used the same models described above and again found that
the high S/N sources (\gtsim 100 counts) did not bias the best-fit
parameters. The best-fit parameters appear in Table~6, while a plot of
$\Gamma$ vs. redshift using these values is shown in
Figure~\ref{gamma_plot}. We found no detectable change in $\Gamma$
with redshift ($\chi^2=0.5$ for 3 degrees of freedom,
$P(\chi^2|\nu)=0.91)$, and basic fitting shows that the maximum
allowed change in $\Gamma$ across this redshift range can be no more
than $\approx 5\%$.  We have also split the sample into higher and
lower optical luminosity halves and performed joint fitting; within
the uncertainties the maximum allowed change of $\Gamma$ with
luminosity can be no greater than $\approx 10\%$.

\begin{deluxetable*}{lccccccccc}
\tablecolumns{10}
%\rotate
\tablenum{6} \tabletypesize{\scriptsize} \tablecaption{Joint
Spectral-Fitting Results\tablenotemark{a}} \tablehead{
\colhead{Sources} & \colhead{Number of} & \colhead{Median} &
\colhead{Median} & \colhead{Total} & \colhead{} &
\colhead{$C$\tablenotemark{b}} & \colhead{} & \colhead{$N_{\rm
H}$\tablenotemark{c}} & \colhead{$C$\tablenotemark{c}} \\
\colhead{Included} & \colhead{Sources} & \colhead{Redshift} &
\colhead{$M_i$} & \colhead{Counts} &
\colhead{$\Gamma$\tablenotemark{b}} & \colhead{Statistic} &
\colhead{$\Gamma$\tablenotemark{c}} & \colhead{($10^{22} {\rm
cm}^{-2}$)} & \colhead{Statistic} } \startdata All RQQs & 42 & 4.04 &
$-29.36$ & 1872 & 1.92$^{+0.09}_{-0.08}$ & 1149.0 &
1.92$^{+0.10}_{-0.06}$ & $<$0.2 & 1149.0 \\ RQQs, $<$100 counts & 37 &
4.07 & $-29.34$ & 993 & 1.93$^{+0.10}_{-0.10}$ & 695.1 &
1.93$^{+0.13}_{-0.10}$ & $<$0.6 & 695.1 \\ RQQs, $>$100 counts & 5 &
2.74 & $-29.86$ & 879& 1.92$^{+0.12}_{-0.09}$ & 454.0 &
1.91$^{+0.16}_{-0.08}$ & $<$0.8 & 454.0 \\ RQQs, $1<z<2$ & 4 & 1.88 &
$-29.41$ & 343 & 1.87$^{+0.19}_{-0.14}$ & 192.9 &
1.91$^{+0.37}_{-0.19}$ & $<$1.2 & 192.9 \\ RQQs, $2<z<3$ & 8 & 2.63 &
$-29.46$ & 784 & 1.95$^{+0.11}_{-0.11}$ & 446.6 &
1.95$^{+0.17}_{-0.11}$ & $<$0.8 & 446.6 \\ RQQs, $3<z<4$ & 8 & 3.37 &
$-29.43$ & 305 & 1.90$^{+0.18}_{-0.18}$ & 213.5 &
1.86$^{+0.19}_{-0.18}$ & $<$1.0 & 213.7 \\ RQQs, $4<z<5$ & 22 & 4.34 &
$-29.17$ & 440 & 1.93$^{+0.16}_{-0.16}$ & 295.7 &
1.87$^{+0.16}_{-0.16}$ & $<$0.8 & 296.5 \\ RQQs, $<$100 counts,
$1<z<2$ & 3 & 1.90 & $-29.41$ & 162 & 1.98$^{+0.25}_{-0.25}$ & 103.9 &
2.13$^{+0.48}_{-0.37}$ & $<$2.0 & 103.9 \\ RQQs, $<$100 counts,
$2<z<3$ & 5 & 2.82 & $-29.39$ & 194 & 1.95$^{+0.23}_{-0.23}$ & 156.3 &
2.10$^{+0.40}_{-0.35}$ & $<$2.5 & 156.3 \\ RQQs, $<$100 counts,
$3<z<4$ & 7 & 3.38 & $-29.41$ & 197 & 1.87$^{+0.23}_{-0.23}$ & 138.9 &
1.87$^{+0.28}_{-0.22}$ & $<$2.1 & 138.9 \\ RQQs, $<$100 counts,
$4<z<5$ & 22 & 4.34 & $-29.17$ & 440 & 1.93$^{+0.16}_{-0.16}$ & 295.7
& 1.87$^{+0.16}_{-0.16}$ & $<$0.8 & 296.5 \\ RQQs, High-Luminosity
Half\tablenotemark{d} & 21 & 3.70 & $-29.57$ & 1151 &
1.94$^{+0.09}_{-0.09}$ & 681.0 & 1.91$^{+0.12}_{-0.08}$ & $<$0.3 &
681.2 \\ RQQs, Low-Luminosity Half\tablenotemark{d} & 21 & 4.03 &
$-29.14$ & 721 & 1.90$^{+0.13}_{-0.11}$ & 468.0 &
1.97$^{+0.18}_{-0.16}$ & $<$1.1 & 468.0 \\ RQQs, $<100$ counts,
High-Lum.\tablenotemark{e} & 18 & 4.09 & $-29.43$ & 508 &
1.93$^{+0.17}_{-0.12}$ & 358.9 & 1.91$^{+0.17}_{-0.12}$ & $<$0.7 &
358.9 \\ RQQs, $<100$ counts, Low-Lum.\tablenotemark{e} & 19 &
4.06 & $-29.02$ & 485 & 1.93$^{+0.15}_{-0.15}$ & 336.3 &
1.98$^{+0.23}_{-0.19}$ & $<$1.3 & 336.3 \\ \enddata
\tablenotetext{a}{Errors on $\Gamma$ and upper limits for $N_{\rm H}$
are quoted at 90\% confidence levels.}  \tablenotetext{b}{Without an
intrinsic absorption component.}  \tablenotetext{c}{With an intrinsic
absorption component.}  \tablenotetext{d}{The high-luminosity and
low-luminosity halves have \xray\ luminsities ranging from $\log (\nu
L_\nu)_{\rm 2~keV}=44.50$--45.76 and $\log (\nu L_\nu)_{\rm
2~keV}=44.45$--45.83, respectively.}  \tablenotetext{e}{The $<100$
count high-luminosity and low-luminosity halves have \xray\
luminsities ranging from $\log (\nu L_\nu)_{\rm 2~keV}=44.50$--45.83
and \hbox{$\log (\nu L_\nu)_{\rm 2~keV}=44.45$--45.48}, respectively.}
\end{deluxetable*}

%%
%% FIGURE 6
%%

At rest-frame energies below $\sim1$~keV, some quasar SEDs can deviate
from a power-law due to the additional \xray\ flux provided by the
soft \xray\ excess (e.g., Porquet et al. 2004). This additional soft
\xray\ emission can bias calculations of $\Gamma$ toward higher
(softer) values. To see if our measurements of $\Gamma$ are biased in
such a way, we redid our joint spectral analysis considering only the
data above rest-frame 2~keV. Since our lower observed-frame energy
limit is still 0.5~keV, this additional constraint does not affect our
calculations at $z\gtsim3$. Implementing this additional constraint
did not change our calculated values of $\Gamma$ significantly. As
perhaps expected, the largest change occurred in the $z=1$--2 bin. In
this redshift bin, the photon index increased from $\Gamma =
1.87^{+0.19}_{-0.14}$ to $\Gamma = 1.95^{+0.22}_{-0.17}$, which is
well within the derived errors (and this increase goes opposite to the
sense expected if soft \xray\ excess emission were present). The small
difference between these two methods shows that our sample is not
significantly biased by excess soft \xray\ emission, so we continue to
use the results from joint fitting done in the observed-frame,
$0.5$--8.0 keV band.

We have searched for a narrow, neutral, iron K$\alpha$ line in each of
the integer redshift-binned sets of spectra. No lines were
detected. Upper limits on the rest-frame EWs of any such emission
appear in Table~7; the rest-frame EWs were calculated at the mean
redshift in each bin. These upper limits range from $\approx 150$--500
eV. These are not particularly tight constraints, especially when
considering the high average luminosity of our sample. Quasars with
higher \xray\ luminosities tend to show weaker iron K$\alpha$ emission
lines (e.g., Page et~al.\ 2004; Bianchi et~al.\ 2007).

\begin{deluxetable}{lcc}
\tablecolumns{3} \tablenum{7} \tabletypesize{\scriptsize}
\tablecaption{Iron K$\alpha$ Line Constraints}
\tablehead{ \colhead{} & \colhead{Rest-Frame} & \colhead{Number of} \\
\colhead{} & \colhead{EW (eV)\tablenotemark{a}} & \colhead{Counts} }
\startdata All RQQs, $1<z<2$ & \ltsim 490.6 & 343 \\ All RQQs, $2<z<3$
& \ltsim 302.9 & 784 \\ All RQQs, $3<z<4$ & \ltsim 462.0 & 305 \\ All
RQQs, $4<z<5$\tablenotemark{b} & \ltsim 144.4 & 440 \\ RQQs, $<100$
counts, $1<z<2$ & \ltsim 883.3 & 162 \\ RQQs, $<100$ counts, $2<z<3$ &
\ltsim 1387.2 & 194 \\ RQQs, $<100$ counts, $3<z<4$ & \ltsim 769.7 &
197 \\ RQQs, $<100$ counts, $4<z<5$\tablenotemark{b} & \ltsim 144.4 &
440 \\ \enddata \tablenotetext{a}{Calculated at the mean redshift in
each bin and quoted at the 90\% confidence level.}
\tablenotetext{b}{Three of the quasars with $<3$ full-band counts were
excluded.}
\end{deluxetable}

We have also checked for a Compton-reflection continuum component at
$\approx10$--50~keV in our spectra, which would be particularly
apparent at high redshifts. No reflection component was found; this is
not unexpected given the high luminosities and relatively low number
of counts for the higher redshift sources.

\subsection{X-ray-to-Optical Spectral Energy Distributions}

\subsubsection{Basic Sample Properties}

The \xray-to-optical flux ratio for AGNs has been found to decrease at
higher optical luminosities, but it does not show any clear change
with redshift (e.g., Avni \& Tananbaum 1986; Wilkes et~al.\ 1994;
Strateva et~al.\ 2005; S06 and references therein; but see Kelly
et~al.\ 2007). Using our 34 object core sample of highly luminous
quasars spanning the widest possible redshift range for such objects
(\hbox{$z\approx 1.5$--4.5}), we further examine the \hbox{\aox-\luv}
relationship and provide constraints on \aox\ evolution with
redshift. All of the statistical analyses presented below have
excluded radio-loud, BAL, gravitationally-lensed, and weak-line
quasars (see below), as well as SDSS~J1521$+$5202 (see \S4); any group
of quasars satisfying these criteria will be hereafter referred to as
``clean''.

\begin{figure*}
%\epsscale{0.8}
\plotone{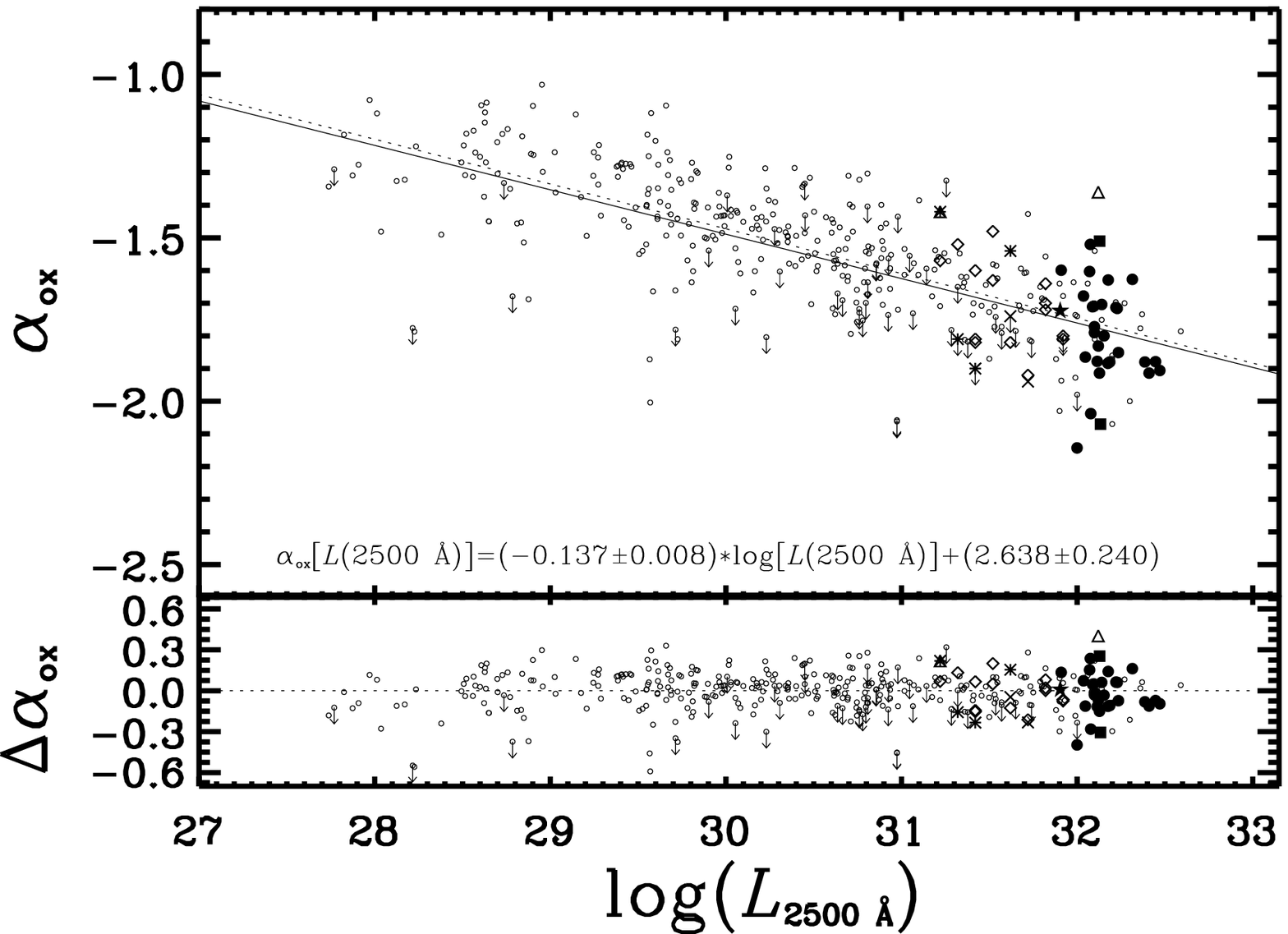} 
\figurenum{7}
\caption{\label{aox_v_lum} \aox\ vs. \luv\ for our core sample (filled
 symbols), the full S06 sample, and the Shemmer et~al. (2006a) sample
 (open diamonds); only clean quasars have been included. Upper limits
 are denoted with downward-pointing arrows; all of our core-sample
 sources have \xray\ detections. The solid (dotted) line is the
 best-fit relation found in this paper (S06). The lower panel shows
 residuals (\daox) from the S06 relation; this equation is shown at
 the bottom of the top panel. The most-luminous object on this plot is
 PSS~0926$+$3055, one of the complementary high-luminosity $z\gtsim4$
 sources.}
\end{figure*}

Figure~\ref{aox_v_lum} shows \aox\ vs. \luv\ for our core sample
combined with the full S06 sample (including the complementary
high-luminosity $z\gtsim4$ quasars) and 14 additional $z>4$ quasars
from Shemmer et~al.\ (2006a), resulting in an \xray\ detection
fraction of 89\%. The inclusion of the 14 clean quasars from Shemmer
et~al.\ (2006a) significantly improves coverage at 
\hbox{$z\approx5$--6}; note that the full Shemmer et~al.\ (2006a) sample
includes four weak emission-line quasars, which we do not include in
our analyses since the nature of these objects remains unclear. The
best-fit relation from S06:
\begin{equation}
\alpha_{\rm ox}=(-0.137\pm 0.008){\rm log}(L_{\mbox{\scriptsize
2500~\AA}})+(2.638 \pm 0.240)
\end{equation} is shown as a dotted line; a more detailed analysis of
the best fit for the correlation between \aox\ and UV luminosity is
given in \S5.2.2 below. Note that the addition of our core sample of
26 clean quasars to the full S06 sample increases by a factor of
\hbox{$\approx 2$} the number of quasars at the highest luminosities
that have \aox\ values.

%%
%% FIGURE 7
%%

\begin{figure}
%\epsscale{0.8}
\plotone{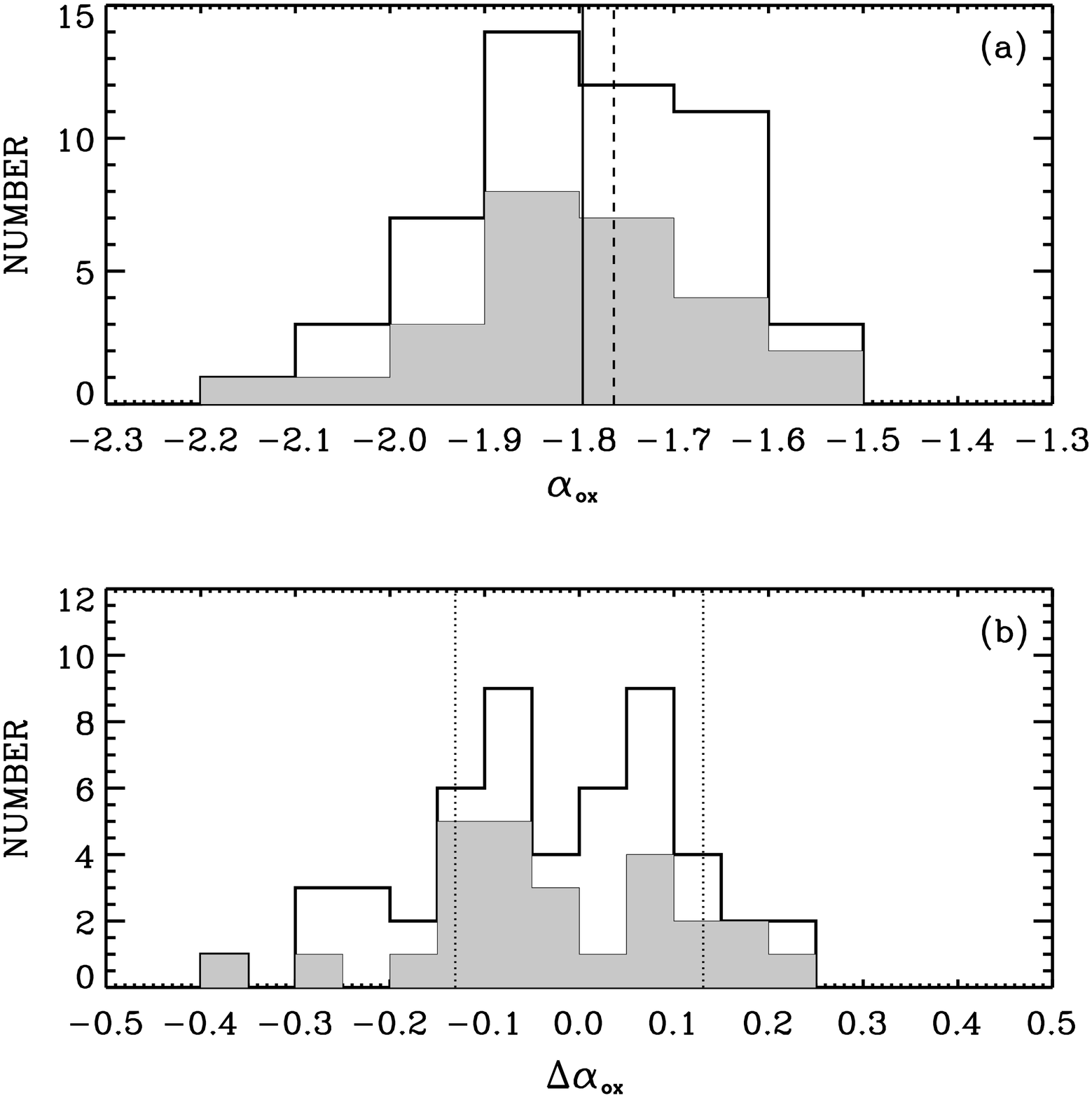} 
\figurenum{8}
\caption{\label{aox_hist} Histograms of ({\it a}) \aox\ and ({\it b})
 \daox\ for our 51 clean-quasar sample, with bin sizes of 0.1 and
 0.05, respectively. Our core sample is marked ({\it light shaded
 histogram}). The solid (dashed) line in panel {\it a} marks the mean
 measured (predicted) \aox\ of our quasar sample. In panel {\it b},
 dotted lines mark the 1~$\sigma$ range ($\sigma = \pm 0.131$) for the
 distribution of \daox\ for sources with \hbox{$32<\log($\luv)~$<33$},
 taken from Table~5 of S06.
}
\end{figure}

In Figure~\ref{aox_hist} we show histograms of our \aox\ distribution
and the distribution of residuals from the S06 best fit (\daox) for
our 51 quasar clean sample. The 26 clean quasars from our 34 quasar
core sample are marked with light shaded histograms, while the 25
complementary $z\gtsim4$ sources from S06 are not marked. Marked on
the \aox\ distribution are the measured and predicted mean \aox\
values, as solid and dashed lines, respectively; the predicted mean
\aox\ value was calculated from equation (2). Some standard
statistical values for our \aox\ and \daox\ distributions are shown in
Table~8. The mean value of \aox~$=-1.80\pm0.02$ for our sample agrees
with the predicted value from Table~5 of S06 ($-1.788$) to within
$1~\sigma$. In the \daox\ histogram, dotted lines mark the 1~$\sigma$
range from the predicted S06 value (see \S3, Column 13). As a test, we
have determined that it would require a reduction of $\approx17\%$ to
the \xray\ luminosities of our sources (corresponding to a change in
\aox\ of $\approx-0.03$) in order for the mean \aox\ of our sample to
become inconsistent with the predicted value of S06. We have also used
the method of Maccacaro et~al.\ (1988) to estimate the {\it intrinsic}
dispersion of the \aox\ values for our 51 non-BAL RQQs. We find a
highly significant intrinsic dispersion of 0.10; the measured
dispersion, not correcting for measurement errors, is 0.14.

\begin{deluxetable*}{lcrccccrc}
\tablecolumns{9} \tablenum{8} \tabletypesize{\scriptsize}
\tablecaption{\aox\ and \daox\ Statistical Values}
\tablehead{ \colhead{} & \colhead{Number of} & \colhead{} &
\colhead{Unweighted} & \colhead{Measured} & \colhead{Intrinsic} &
\colhead{1st} & \colhead{} & \colhead{3rd} \\ \colhead{} &
\colhead{Sources} & \colhead{} & \colhead{Mean} & \colhead{Dispersion}
& \colhead{Dispersion} & \colhead{Quartile} & \colhead{Median} &
\colhead{Quartile} } \startdata All Objects & 51 & \aox\ & $-1.797\pm
0.019$ & 0.136 & 0.103 & $-1.880$ & $-1.800$ & $-1.700$ \\ & & \daox\
& $-0.029\pm 0.019$ &\nodata&\nodata& $-0.110$ & $-0.033$ & $+0.072$
\\ $1<z<2$ & 4 & \aox\ & $-1.805\pm 0.060$ & 0.120
&\nodata\tablenotemark{a}& $-1.880$ & $-1.878$ & $-1.865$ \\ & &
\daox\ & $-0.044\pm 0.056$ &\nodata&\nodata& $-0.111$ & $-0.110$ &
$-0.106$ \\ $2<z<3$ & 11 & \aox\ & $-1.832\pm 0.041$ & 0.136 & 0.093 &
$-1.942$ & $-1.841$ & $-1.758$ \\ & & \daox\ & $-0.057\pm 0.041$
&\nodata&\nodata& $-0.120$ & $-0.063$ & $+0.009$ \\ $3<z<4$ & 11 &
\aox\ & $-1.765\pm 0.045$ & 0.149 & 0.125 & $-1.880$ & $-1.713$ &
$-1.678$ \\ & & \daox\ & $+0.010\pm 0.047$ &\nodata&\nodata& $-0.079$
& $+0.066$ & $+0.084$ \\ $4<z<5$ & 25 & \aox\ & $-1.795\pm 0.026$ &
0.130 & 0.070 & $-1.871$ & $-1.796$ & $-1.709$ \\ & & \daox\ &
$-0.035\pm 0.026$ &\nodata&\nodata& $-0.130$ & $-0.022$ & $+0.061$ \\
\enddata \tablenotetext{a}{Given the low number of sources in this
redshift bin, we were unable to accurately determine the intrinsic
dispersion.}
\end{deluxetable*}

As a basic first test for any redshift dependence of \aox, we have
binned our sample of quasars, including the complementary
high-luminosity $z\gtsim4$ sources, into integer redshift bins from
$z=1$--5. Recall from \S5.1.2 that the mean \luv\ in each bin does not
differ from the sample mean by more than $2~\sigma$, with three of the
four redshift bins differing by less than $1~\sigma$; this reduces the
effect of the \aox-luminosity correlation when looking for any
\aox-$z$ correlation. The values for the mean \aox\ calculated in each
bin, as well as the mean residuals (i.e., \daox) from the S06 best fit
and the best fit found in this paper [see \S5.2.2, equation (3)], are
plotted against redshift in Figure~\ref{aox_v_z}. Error bars show the
standard error of the mean in each bin. No detectable change in \aox\
is evident across the full redshift range [a constant fit gives
$\chi^2=1.3$, $P(\chi^2|\nu)=0.74$], and using basic fitting we have
placed a constraint on any such change in \aox\ to be $\ltsim 6\%$
(corresponding to a change in the ratio of \lx\ to \luv\ of less than
a factor of 1.9).

%%
%% FIGURE 8
%%

\subsubsection{Linear-Regression Analysis}

To investigate further correlations between \aox, \luv, \lx, and $z$,
we have added our core sample of 26 clean quasars to the full 333
source sample of S06; we removed SDSS~J1701$+$6412 from the S06 sample
since it is present in both. Also included are the 14 $z>4$ clean
quasars from Shemmer et~al.\ (2006a). The inclusion of our core sample
of quasars, which lie in a narrow range of high luminosity
[log(\luv)~$\approx 32.0$--32.5] and across a fairly wide range of
redshift ($z\approx 1.5$--4.5), allows exploration of a new region of
the luminosity-redshift plane (see Figure~\ref{lum_v_z}). Also shown
in Figure~\ref{lum_v_z} are the additional quasars from Shemmer
et~al.\ (2006a), which substantially improve coverage at
$z\approx5$--6. Ultimately, 372 quasars are included in our analysis:
26 from our core sample, 332 of the 333 from the full S06 sample, and
14 from the Shemmer et~al.\ (2006a) sample, increasing the S06 sample
size by $\approx12\%$. Note that we do not expect significant
problematic effects from unidentified BAL quasars at $z\ltsim1.5$ in
the S06 sample (see \S3.3 of Strateva et~al.\ 2005).

%%
%% FIGURE 9
%%
\begin{figure}
%\epsscale{0.6}
\plotone{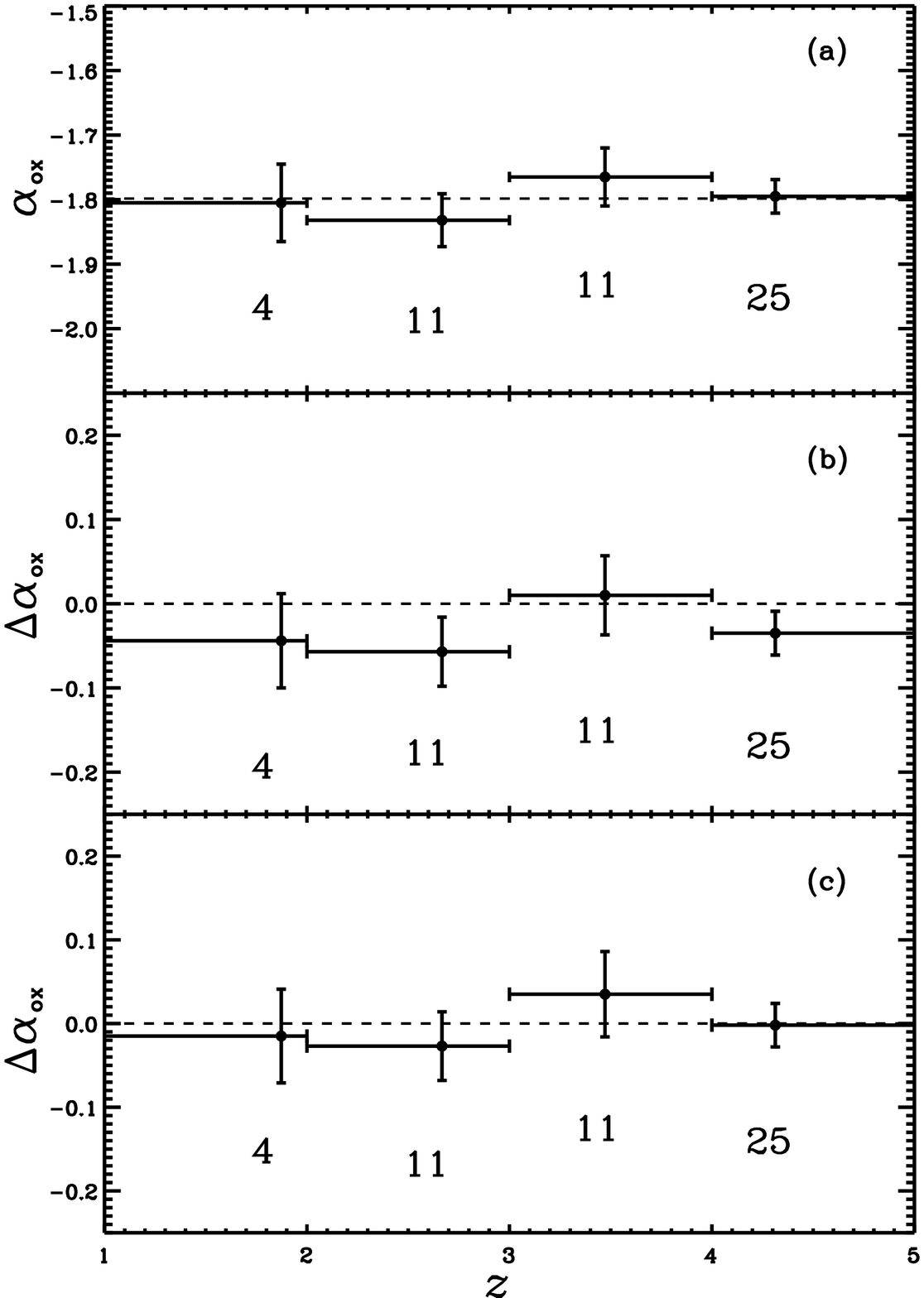} 
\figurenum{9}
\caption{\label{aox_v_z} ({\it a}) \aox\ plotted against redshift,
 binned into four integer redshift bins ranging from $z=1$--5. The
 complementary high-luminosity sources at $z\gtsim4$ from S06 (see
 \S2.1.2) are also included in this analysis. The number of quasars in
 each bin is given below the error bars; these numbers differ from
 those in Figure 6 due to the inclusion of \rosat\ and \xmm\ targets,
 as well as the three sources with $<2$~counts (SDSS~1350$+$5716,
 BR~1117$-$1329 and PSS~2344$+$0342). The vertical error bars show the
 standard error of the mean in each bin, while the horizontal error
 bars show the width of each bin; the data points are marked at the
 median redshift in each bin. In panel ({\it b}) are the residuals
 from the best-fit relation derived in S06, while in panel ({\it c})
 are residuals from Equation (3) of this paper, i.e. the best-fit
 relation found in our study. The best constant-model fit (giving
 \aox~$=-1.80$) is shown as a dashed line in panel ({\it a}).  }
\end{figure}

To quantify the correlations found between the optical and \xray\
properties, we used the Astronomy Survival Analysis software package
(ASURV rev 1.2; Isobe et~al.\ 1990; Lavalley et~al.\ 1992) to perform
linear regressions on the data. ASURV treats censored data using the
survival-analysis methods presented in Feigelson \& Nelson (1985) and
Isobe et~al.\ (1986). We used both the fully parametric EM (estimate
and maximize) regression algorithm (Dempster et~al.\ 1977) and the
semiparametric Buckley-James regression algorithm (Buckley \& James
1979) when performing linear regressions. In what follows we report
the parameters derived from the EM regression, although in all cases
the Buckley-James regression algorithm agreed within the errors.

We confirm and strengthen the finding in previous studies that \aox\
decreases with increasing rest-frame UV luminosity. Performing linear
regressions with ASURV on the combined sample of 372 quasars, we found
the best-fit relation between \aox\ and \luv\ to be
\begin{equation}
\alpha_{\rm ox}=(-0.140\pm 0.007){\rm log}(L_{\mbox{\scriptsize
2500~\AA}})+(2.705 \pm 0.212).
\end{equation}
For comparison, both our best fit as well as the S06 best fit are
shown in Figure~\ref{aox_v_lum} as solid and dashed lines,
respectively.

We also confirm a significant correlation exists between \aox\ and \lx. The
best-fit parameters for this relation are
\begin{equation}
\alpha_{\rm ox}=(-0.093\pm 0.014){\rm log}(L_{\rm 2~keV})+(0.899 \pm
0.359).
\end{equation}
Note that the EM and Buckley-James regression algorithms are no longer
strictly valid when double-censoring is present (upper limits exist on
both \aox\ and \lx\ in the S06 data). However, given the high \xray\
detection fraction of our combined sample (89\%), we have treated the
censored \lx\ data as though they were detected.

Studies measuring a relationship between \xray\ and UV luminosities of
the form \hbox{$L_{\rm X}\propto L^\beta_{\rm UV}$} have found both
$\beta \simeq 0.7$--0.8 (e.g., Avni \& Tananbaum 1982, 1986; Kriss \&
Canizares 1985; Anderson \& Margon 1987; Wilkes et~al. 1994; Vignali
et~al. 2003; Strateva et~al.\ 2005; S06) and $\beta = 1$ (La Franca
et~al. 1995). We find the best-fit parameters for the \lx-\luv\
relation to be
\begin{equation}
{\rm log}(L_{\rm 2~keV})=(0.636\pm 0.018){\rm log}(\luv)+(7.055 \pm
0.553)
\end{equation}
while treating \lx\ as the dependent variable, and
\begin{equation}
{\rm log}(L_{\rm 2~keV})=(0.808\pm 0.021){\rm log}(\luv)+(1.847 \pm
0.694)
\end{equation}
while treating \lx\ as the independent variable. Using the equations
given in Table~1 of Isobe et~al.\ (1990), we calculate the bisector of
the two lines to be
\begin{equation}
{\rm log}(L_{\rm 2~keV})=(0.709\pm 0.010){\rm log}(\luv)+(4.822 \pm
0.627).
\end{equation}
This result agrees with those previous studies which found $\beta$ to
be inconsistent with unity.

Finally, we used ASURV to investigate the relationship between \aox,
\luv, and $z$. We tested three different parametric forms of redshift
dependence: (1) a dependence on $z$, (2) a dependence on log(1+$z$),
and (3) a dependence on the cosmological look-back time, $\tau(z)$, in
units of the present age of the Universe. The best-fit parameters for
these three relations are
\begin{eqnarray}
\alpha_{\rm ox} & = & (-0.134\pm0.011){\rm log}(\luv) \nonumber \\
 & & - (0.005\pm0.007)z+(2.543\pm 0.320)
\end{eqnarray}
\begin{eqnarray}
\alpha_{\rm ox} & = & (-0.137\pm0.012){\rm log}(\luv) \nonumber \\
 & & -(0.006\pm0.023){\rm log}(1+z)+(2.635\pm 0.340)
\end{eqnarray}
\begin{eqnarray}
\alpha_{\rm ox} & = & (-0.143\pm0.011){\rm log}(\luv) \nonumber \\
 & & -(0.001\pm0.003)\tau(z)+(2.824\pm 0.350).
\end{eqnarray}
All three parametrizations have redshift-dependent coefficients
consistent with zero; note that these equations have the same
parametric form as equations (8)--(10) of Kelly et~al.\ (2007). This
finding agrees with previous studies that have found no evolution of
\aox\ with redshift (e.g., Strateva et~al.\ 2005, S06), as well as our
results from \S5.2.1 (also see Figure~\ref{aox_v_z}).

%%
%% FIGURE 10
%%
\begin{figure}
%\epsscale{0.8}
\plotone{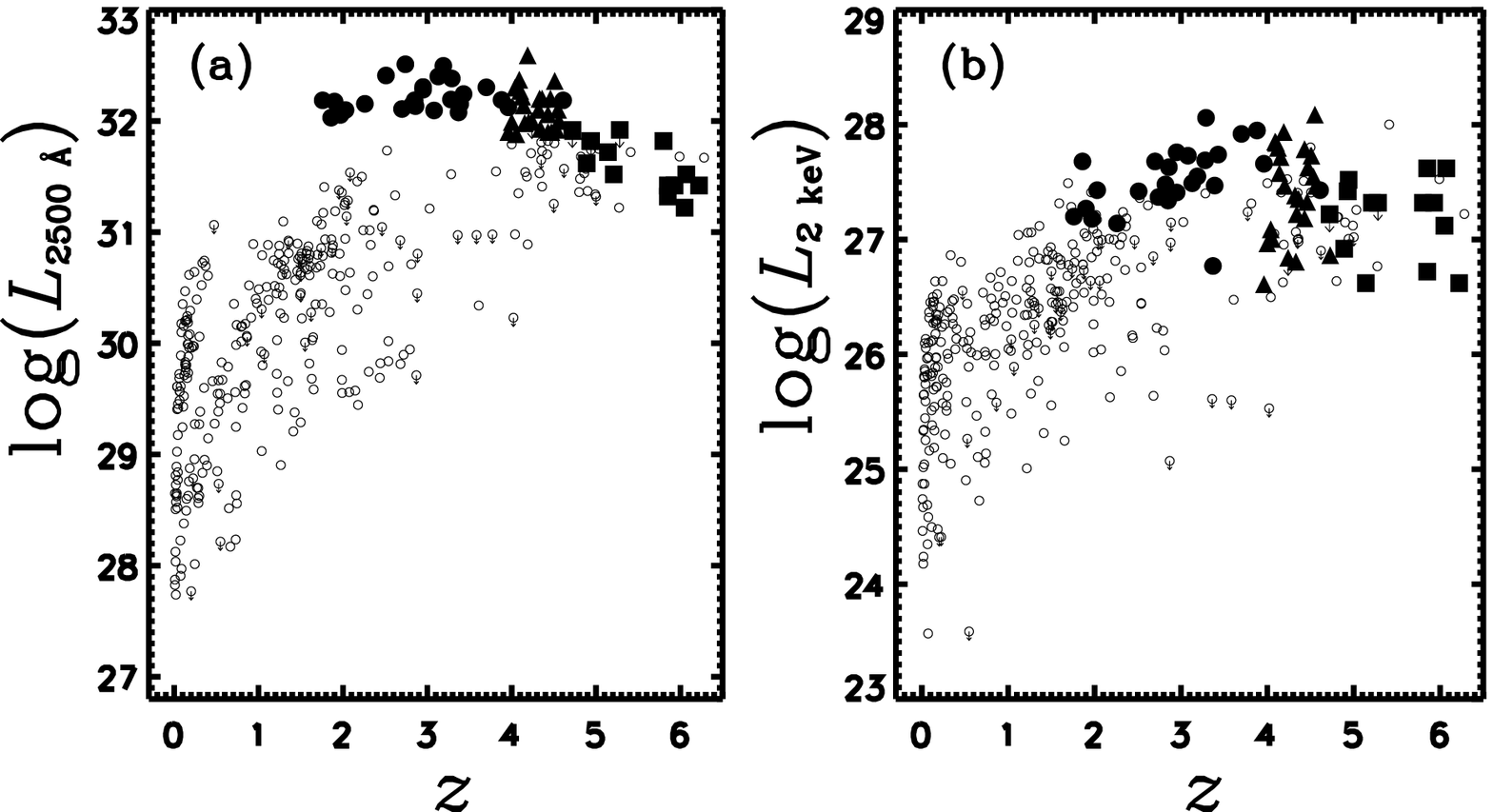}
\figurenum{10}
\caption{\label{lum_v_z} ({\it a}) \luv\ and ({\it b}) \lx\
 vs. redshift for our core sample (26 filled circles) and the
 complementary $z\gtsim4$ sources from S06 (25 filled triangles). Also
 shown are the remaining S06 sample (307 open circles) and the Shemmer
 et~al. (2006a) sample (14 filled squares). Gravitationally-lensed,
 radio-loud, and BAL quasars have been removed, as well as the extreme
 outlier SDSS~J1521$+$5202 (see \S4). Note the new region of
 luminosity-redshift space being populated by our core sample. }
\end{figure}

\begin{figure}
%\epsscale{0.8}
\figurenum{11}
\plotone{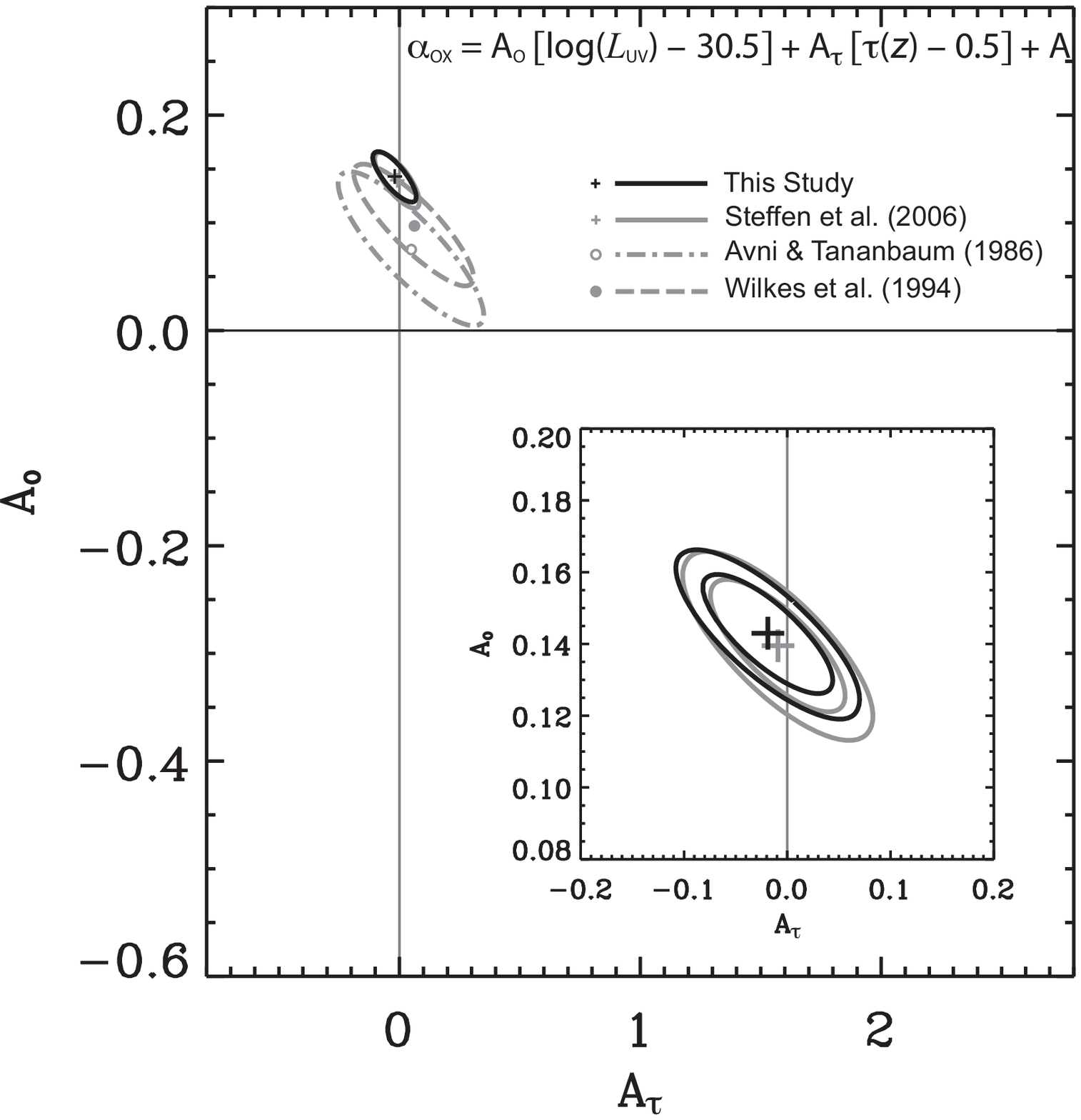} 
\caption{\label{coeff_contours} Best-fit values and 90\% confidence
 contours for the coefficients $A_O$ and $A_\tau$ for our sample of 59
 quasars combined with the S06 sample and the Shemmer et~al. (2006a)
 sample ({\it cross, dark solid contour}), a total of 372 clean
 quasars. Also shown are best-fit values and contours for the S06
 sample ({\it cross, light solid contour}), the sample of Avni \&
 Tananbaum (1986; {\it open circle, dot-dashed contour}), and the
 sample of Wilkes et~al. (1994; {\it filled circle, dashed
 contour}). {\it Inset:} Magnified view of the 68\% and 90\% contours
 for both our sample ({\it solid contours}) and the S06 sample ({\it
 light solid contours}).  }
\end{figure}

We have followed the method described in \S4.4 of S06 to compare our
results directly with those from earlier studies, in particular Avni
\& Tananbaum (1986), Wilkes et~al. (1994), and S06. Confidence
contours of $A_O$ [the coefficient of log(\luv)] and $A_\tau$ [the
coefficient of $\tau(z)$] were calculated using the method outlined in
\S\S 3 and 4 of Avni \& Tananbaum (1986). Renaming $A_z$ as $A_\tau$,
equation (6) of Avni \& Tananbaum (1986) becomes
\begin{eqnarray}
\bar{\alpha}_{\rm ox}(\luv, z | \xray\ {\rm loud}) & = & A_O[{\rm
 log}(\luv)-30.5] \nonumber \\ & & +A_\tau[\tau(z)-0.5]+A.
\end{eqnarray}
We found best-fit values of $[A_O, A_\tau,
A]=[0.143\pm0.011,-0.016\pm0.041,1.556\pm0.009]$ for our
sample. Figure~\ref{coeff_contours} shows contours of $A_O$ and
$A_\tau$ at both the 68\% and 90\% confidence levels, along with
contours from previous studies. Our best-fit value has smaller
confidence contours than those of previous studies and has an $A_\tau$
consistent with zero.

Some Monte Carlo simulations have suggested that correlations among
\aox, \luv, and \lx\ may arise from the effects of luminosity
dispersion in optically selected, flux-limited samples (e.g., Yuan
et~al.\ 1998; Tang et~al.\ 2007). However, these studies have usually
examined the effects of luminosity dispersion over a much smaller
total range in UV luminosity (\hbox{$\Delta$log\luv$\sim 2.5$}) than
our full sample covers (\hbox{$\Delta$log\luv$\sim 5$}). In \S3.5 of
Strateva et~al.\ (2005) the authors estimated the dispersions of \luv\
and \lx\ ($\sigma_{\rm UV}$ and $\sigma_{\rm X}$, when expressed in
log units) and, using simulations, showed that the dispersions cannot
be responsible for the non-unity \lx$-$\luv\ slope they found. They
estimated that $\sigma_{\rm UV}/\sigma_{\rm X}$ is not larger than 1.4
and is plausibly $<1$ for their sample, and we expect these values to
hold for our sample as well. The strength of our sample (which builds
upon Strateva et~al.\ 2005) comes, in part, from the large luminosity
range we cover, which is much larger than the value of the luminosity
dispersion in either band. Tang et~al.\ (2007), in their \S3, do not
%
%%
%% FIGURE 11
%%
take full advantage of the large luminosity range and examine only a
subsample from S06 [and the smaller, high-redshift sample from Miyaji
et~al.\ (2006)], where the luminosity range covered is only slightly
larger than the dispersion.  Comparison of the parametric fits
calculated by Tang et~al.\ (2007, see their Figure 8) with our full
sample shows very significant disagreement between the data and the
fits, especially at high luminosities, where our sample of the
most-luminous quasars helps considerably. Tang et~al.\ (2007), in
their \S5, consider selection effects using Monte Carlo simulations of
a sample more similar to the full S06 sample. They assume the true
slope of the \lx$-$\luv\ relation is unity and assess how much the
observed slope, $\beta$, can differ from unity. They require
unrealistic combinations of $\sigma_{\rm UV}$ and $\sigma_{\rm X}$
(e.g., with $\sigma_{\rm UV} / \sigma_{\rm X} \gg 1$) in order to
obtain values of $\beta$ as flat as the 0.709 that we measure in
equation (7). While our $\beta$ value may be biased somewhat downward
owing to inevitable selection effects in the presently available
samples, it seems unlikely that the true slope of the \lx$-$\luv\
relation could be unity for optically selected quasar samples.

\section{Conclusions and Summary}

We have analyzed the \xray\ properties of a large and statistically
representative sample of the most optically luminous quasars spanning
$M_i\approx -29.3$ to $-30.2$ across a redshift range of
$z\approx1.5$--4.5. Our total quasar sample consists of 59 sources,
including 32 from the SDSS, two additional quasars that were missed by
the SDSS selection criteria, and 25 comparably luminous $z\gtsim4$
quasars. All of these sources have sensitive \xray\ coverage from
either targeted \chandra\ observations or from archival \chandra,
\rosat, or \xmm\ observations; 58 of the 59 sources (98\%) have \xray\
detections. For some of our analyses we have included 332 quasars from
S06 and 14 quasars from Shemmer et~al.\ (2006a). Our main results,
derived for radio-quiet, non-BAL quasars that are not gravitationally
lensed (i.e., our ``clean'' quasars), are the following:
\begin{itemize}
\item{The mean \xray\ power-law photon index for our sample of the
most-luminous quasars is \hbox{$\Gamma=1.92^{+0.09}_{-0.08}$},
consistent with values found in earlier studies.}
\item{Any intrinsic absorbing material for the most-luminous quasars
has been constrained to have a mean column density of $N_{\rm
H}\ltsim2 \times 10^{21}$~${\rm cm}^{-2}$, showing that the
most-luminous quasars typically have little intrinsic \xray\
absorption.}
\item{Using joint spectral fitting, we have found no significant
change in $\Gamma$ with cosmic time over the redshift range
$z\approx1.5$--4.5.}
\item{The mean \aox\ value of the most-luminous quasars is
\aox~$=-1.80\pm0.02$; this is in agreement with earlier studies and
agrees with the predicted value from S06 [at a mean luminosity of
\hbox{log(\luv)~$=32.2$}] to within $1~\sigma$.}
\item{In our sample there is no significant change in \aox\ with
redshift from $z\approx1.5$--4.5 when binned and compared to a
constant model, consistent with results found in some earlier
studies.}
\item{Combining our sample with that of S06 and Shemmer et~al.\
(2006a) results in a clean sample of 372 quasars. Using a parametric
modelling method on this sample, we found \aox\ is clearly dependent
on \luv, but shows no significant dependence on redshift (for three
different parameterizations of redshift dependence). The
\xray-to-optical flux ratios of quasars have not significantly evolved
out to $z\sim6$, and in particular have not significantly evolved out
to $z\sim4.5$ for the most-luminous quasars.}
\end{itemize}

\acknowledgements We gratefully acknowledge the financial support of
NASA grant SAO SV4-74018 (GPG, Principal Investigator), NASA LTSA
grant NAG5-13035 (DWJ, WNB, OS, ATS, DPS), Chandra X-ray Center grant
GO5-6094X (DWJ, WNB, ATS), and NSF grants AST-0607634 and AST-0307582
(DPS).
We thank E.~O. Ofek, A.~W. Rengstorf, and G.~T. Richards for helpful
discussions.
Funding for the creation and distribution of the SDSS Archive has been
provided by the Alfred P. Sloan Foundation, the Participating
Institutions, the National Aeronautics and Space Administration, the
National Science Foundation, the US Department of Energy, the Japanese
Monbukagakusho, and the Max Planck Society. The SDSS Web site is
http://www.sdss.org.
The HET is a joint project of the University of Texas at Austin, the
Pennsylvania State University, Stanford University,
Ludwig-Maximillians-Universitat Munchen, and Georg-August-Universitat
Gottingen. The HET is named in honor of its principal benefactors,
William P. Hobby and Robert E. Eberly.

%{\it Facilities:} \facility{CXO (ACIS)}, \facility{HST (ACS)},
%\facility{Max Plank:2.2m (WFI)}, \facility{VLA}, \facility{ROSAT
%  (PSPC)}, \facility{Sloan}

%\bibliographystyle{aj}
%\bibliography{ms}

%\bibliographystyle{proposal}
%\bibliography{references}

\setcounter{table}{3}

\clearpage
\begin{landscape}
\begin{deluxetable}{lcccrcrrrrcccr}
\tablecolumns{14}
%\rotate
%\center
\tablenum{4} \tabletypesize{\tiny} \tablecaption{X-ray, Optical, and
Radio Properties of the Core Sample} \tablehead{ \colhead{} &
\colhead{} & \colhead{} & \colhead{} & \colhead{} &
\colhead{$\log(L_\nu)$} & \colhead{Count} & \colhead{} & \colhead{} &
\colhead{$\log (\nu L_\nu)$} & \colhead{$\log L$} & \colhead{} &
\colhead{} & \colhead{} \\ \colhead{Object (SDSS~J)} &
\colhead{$N_{\rm H}$} & \colhead{$AB_{1450}$} & \colhead{$M_i$} &
\colhead{$f_{2500~{\rm \AA}}$\tablenotemark{a}} & \colhead{2500~\AA} &
\colhead{Rate\tablenotemark{b}} & \colhead{$f_{\rm
x}$\tablenotemark{c}} & \colhead{$f_{2~\rm keV}$\tablenotemark{d}} &
\colhead{${2~\rm keV}$} & \colhead{$2$--10~keV} &
\colhead{$\alpha_{\rm{ox}}$} & \colhead{${\Delta}\alpha_{\rm{ox}}$
($\sigma$)\tablenotemark{e}} & \colhead{$R$} \\ \colhead{(1)} &
\colhead{(2)} & \colhead{(3)} & \colhead{(4)} & \colhead{(5)} &
\colhead{(6)} & \colhead{(7)} & \colhead{(8)} & \colhead{(9)} &
\colhead{(10)} & \colhead{(11)} & \colhead{(12)} & \colhead{(13)} &
\colhead{(14)} } \startdata 012156.04$+$144823.9 & 3.92 & 17.0 &
$-$29.29 & 6.83 & 32.09 & 9.9$^{+1.9}_{-1.6}$ & 41.2$^{+7.7}_{-6.6}$ &
23.8$^{+4.5}_{-3.8}$ & 45.31$^{+0.07}_{-0.08}$ & 45.52 & $-$1.71 &
$+$0.05 (0.36) & $<$ 2.9\tablenotemark{f} \\ 014516.59$-$094517.3 &
2.70 & 16.7 & $-$29.50 & 9.08 & 32.18 & 42.5$^{+2.0}_{-1.8}$ &
189.4$^{+8.4}_{-7.9}$ & 105.5$^{+4.7}_{-4.4}$ &
45.93$^{+0.02}_{-0.02}$ & 46.13 & $-$1.51 & $+$0.20
(1.34)\tablenotemark{g} & $<$ 0.4 \\ 020950.71$-$000506.4 & 2.42 &
16.9 & $-$29.40 & 7.44 & 32.12 & 5.5$^{+2.1}_{-1.6}$ &
21.9$^{+8.4}_{-6.2}$ & 12.6$^{+4.8}_{-3.6}$ & 45.03$^{+0.14}_{-0.15}$
& 45.24 & $-$1.83 & $-$0.07 (0.52) & $<$ 0.5 \\ 073502.31$+$265911.4 &
5.67 & 16.5 & $-$29.28 & 11.85 & 32.05 & 8.5$^{+1.7}_{-1.5}$ &
37.0$^{+7.5}_{-6.3}$ & 16.4$^{+3.4}_{-2.8}$ & 44.87$^{+0.08}_{-0.08}$
& 45.08 & $-$1.87 & $-$0.11 (0.86) & 1.0 \\ 075054.64$+$425219.2 &
4.95 & 16.0 & $-$29.50 & 17.08 & 32.17 & 11.4$^{+1.9}_{-1.7}$ &
48.5$^{+8.3}_{-7.1}$ & 21.0$^{+3.6}_{-3.1}$ & 44.95$^{+0.07}_{-0.07}$
& 45.16 & $-$1.88 & $-$0.11 (0.87) & $<$ 0.2 \\ 080342.04$+$302254.6 &
4.55 & 16.3 & $-$29.33 & 12.68 & 32.10 & 14.4$^{+2.2}_{-1.9}$ &
60.7$^{+9.1}_{-8.0}$ & 27.5$^{+4.1}_{-3.6}$ & 45.12$^{+0.06}_{-0.06}$
& 45.32 & $-$1.79 & $-$0.03 (0.23) & $<$ 0.3 \\ 081331.28$+$254503.0 &
3.80 & 16.2 & $-$29.40 & 13.45 & 31.89 & 88.7$^{+4.5}_{-4.3}$ &
281.3$^{+14.2}_{-13.5}$ & 105.5$^{+5.3}_{-5.1}$ &
45.47$^{+0.02}_{-0.02}$ & 45.68 & $-$1.58 & $+$0.06
(0.39)\tablenotemark{g} & $<$ 0.2 \\ 084401.95$+$050357.9 & 3.65 &
17.7 & $-$29.49 & 3.46 & 31.90 & 4.2$^{+1.3}_{-1.1}$ &
17.3$^{+5.5}_{-4.3}$ & 11.2$^{+3.6}_{-2.8}$ & 45.10$^{+0.12}_{-0.12}$
& 45.31 & $-$1.72 & $+$0.01 (0.07) & 18.9 \\ 090033.49$+$421546.8 &
2.03 & 16.6 & $-$29.86 & 9.17 & 32.31 & 21.0$^{+2.6}_{-2.3}$ &
82.5$^{+10.2}_{-9.1}$ & 52.8$^{+6.5}_{-5.8}$ & 45.76$^{+0.05}_{-0.05}$
& 45.97 & $-$1.63 & $+$0.16 (1.24) & 1.6 \\ 094202.04$+$042244.5 &
3.51 & 17.1 & $-$29.39 & 6.18 & 32.14 & 8.6$^{+1.7}_{-1.5}$ &
35.1$^{+7.1}_{-5.9}$ & 22.4$^{+4.5}_{-3.8}$ & 45.38$^{+0.08}_{-0.08}$
& 45.59 & $-$1.70 & $+$0.06 (0.47) & $<$ 0.6 \\ 095014.05$+$580136.5 &
1.35 & 17.6 & $-$29.28 & 3.59 & 32.04 & 5.3$^{+1.4}_{-1.2}$ &
20.6$^{+5.5}_{-4.4}$ & 15.2$^{+4.1}_{-3.3}$ & 45.35$^{+0.10}_{-0.11}$
& 45.55 & $-$1.68 & $+$0.07 (0.56) & $<$ 1.3 \\ 100129.64$+$545438.0 &
0.84 & 15.9 & $-$29.41 & 17.06 & 32.11 & 14.0$^{+2.1}_{-1.9}$ &
52.9$^{+8.1}_{-7.1}$ & 21.8$^{+3.3}_{-2.9}$ & 44.91$^{+0.06}_{-0.06}$
& 45.11 & $-$1.88 & $-$0.12 (0.89) & $<$ 0.2 \\ 101447.18$+$430030.1 &
1.16 & 16.5 & $-$30.02 & 11.76 & 32.38 & 6.3$^{+1.5}_{-1.2}$ &
24.1$^{+5.8}_{-4.7}$ & 14.9$^{+3.5}_{-2.9}$ & 45.17$^{+0.09}_{-0.10}$
& 45.38 & $-$1.88 & $-$0.08 (0.62) & $<$ 0.3 \\ 110610.73$+$640009.6 &
1.11 & 16.1 & $-$29.65 & 14.83 & 32.23 & 27.4$^{+3.0}_{-2.7}$ &
104.7$^{+11.6}_{-10.4}$ & 50.0$^{+5.6}_{-5.0}$ &
45.44$^{+0.05}_{-0.05}$ & 45.65 & $-$1.72 & $+$0.06 (0.47) & $<$ 0.2
\\ 111038.64$+$483115.7 & 1.37 & 16.7 & $-$29.76 & 9.11 & 32.23 &
\nodata & 24.0$^{+0.4}_{-0.4}$ & 13.7$^{+0.2}_{-0.2}$ &
45.10$^{+0.01}_{-0.01}$ & 45.30 & $-$1.85 & $-$0.07 (0.56) & $<$ 0.4
\\ 121930.78$+$494052.3 & 1.83 & 17.0 & $-$29.29 & 7.27 & 32.07 &
7.2$^{+2.3}_{-1.8}$ & 87.8$^{+28.4}_{-22.1}$ & 48.5$^{+15.7}_{-12.2}$
& 45.58$^{+0.12}_{-0.13}$ & 45.79 & $-$1.60 & $+$0.15 (1.17) & 2.4 \\
123549.47$+$591027.0 & 1.18 & 16.9 & $-$29.46 & 7.18 & 32.10 &
8.0$^{+1.7}_{-1.4}$ & 30.4$^{+6.5}_{-5.5}$ & 17.4$^{+3.7}_{-3.1}$ &
45.17$^{+0.08}_{-0.09}$ & 45.37 & $-$1.77 & $-$0.01 (0.10) & $<$ 0.5
\\ 123641.46$+$655442.0 & 1.96 & 17.2 & $-$29.43 & 6.03 & 32.15&
4.8$^{+1.4}_{-1.1}$ & 18.8$^{+5.4}_{-4.3}$ & 12.3$^{+3.5}_{-2.8}$ &
45.15$^{+0.11}_{-0.11}$ & 45.35 & $-$1.80 & $-$0.03 (0.25) & $<$
3.7\tablenotemark{f} \\ 135044.67$+$571642.8 & 1.22 & 17.2 & $-$29.31
& 5.45 & 32.00 & 0.6$^{+0.9}_{-0.4}$ & 2.5$^{+3.4}_{-1.7}$ &
1.4$^{+2.0}_{-1.0}$ & 44.10$^{+0.37}_{-0.48}$ & 44.31 & $-$2.14 &
$-$0.40 (3.02) & $<$ 0.7 \\ 140747.22$+$645419.9 & 1.90 & 17.2 &
$-$29.29 & 6.16 & 32.09 & 11.0$^{+2.2}_{-1.9}$ & 43.0$^{+8.7}_{-7.4}$
& 26.2$^{+5.3}_{-4.5}$ & 45.41$^{+0.08}_{-0.08}$ & 45.61 & $-$1.68 &
$+$0.08 (0.62) & $<$ 3.4\tablenotemark{f} \\ 142123.98$+$463317.8 &
1.40 & 17.3 & $-$29.34 & 5.12 & 32.08 & 1.0$^{+0.8}_{-0.5}$ &
3.8$^{+3.1}_{-1.9}$ & 2.5$^{+2.0}_{-1.2}$ & 44.45$^{+0.26}_{-0.29}$ &
44.66 & $-$2.04 & $-$0.28 (2.15) & $<$ 0.7 \\ 142656.18$+$602550.9 &
1.75 & 16.3 & $-$30.22 & 13.05 & 32.45 & 2.2$^{+1.0}_{-0.7}$ &
26.5$^{+12.2}_{-8.8}$ & 16.6$^{+7.7}_{-5.5}$ & 45.23$^{+0.17}_{-0.17}$
& 45.44 & $-$1.88 & $-$0.07 (0.55) & $<$ 0.3 \\ 143835.95$+$431459.2 &
1.61 & 17.6 & $-$29.60 & 4.00 & 32.18 & 2.1$^{+1.1}_{-0.7}$ &
8.2$^{+4.1}_{-2.9}$ & 6.9$^{+3.5}_{-2.4}$ & 45.11$^{+0.18}_{-0.19}$ &
45.31 & $-$1.83 & $-$0.06 (0.44) & $<$ 1.1 \\ 144542.76$+$490248.9 &
2.27 & 17.4 & $-$29.41 & 4.06 & 32.07 & 4.9$^{+1.1}_{-0.9}$ &
61.1$^{+13.9}_{-11.5}$ & 44.5$^{+10.1}_{-8.4}$ &
45.80$^{+0.09}_{-0.09}$ & 46.01 & $-$1.52 & $+$0.24 (1.80) & 7.6 \\
152156.48$+$520238.4 & 1.59 & 15.8 & $-$30.17 & 21.34 & 32.38 &
0.7$^{+0.7}_{-0.4}$ & 2.1$^{+2.2}_{-1.2}$ & 1.0$^{+1.1}_{-0.6}$ &
43.74$^{+0.32}_{-0.38}$ & 43.95 & $-$2.44 & $-$0.63 (4.82) & $<$ 0.1
\\ 152553.89$+$513649.1 & 1.60 & 16.9 & $-$29.64 & 7.41 & 32.13 &
\nodata & 129.2$^{+12.1}_{-13.8}$ & 74.9$^{+7.0}_{-8.0}$ &
45.82$^{+0.03}_{-0.05}$ & 45.92 & $-$1.51 & $+$0.25 (1.93) & $<$ 0.5
\\ 161434.67$+$470420.0 & 1.20 & 16.4 & $-$29.36 & 9.59 & 31.91 &
49.1$^{+4.8}_{-4.4}$ & 153.3$^{+14.9}_{-13.7}$ & 65.5$^{+6.4}_{-5.8}$
& 45.43$^{+0.04}_{-0.04}$ & 45.63 & $-$1.60 & $+$0.13 (0.92) & 4.8 \\
162116.92$-$004250.8 & 7.11 & 17.0 & $-$29.69 & 5.53 & 32.18 &
12.7$^{+3.5}_{-2.8}$ & 44.8$^{+12.4}_{-9.9}$ & 31.4$^{+8.7}_{-7.0}$ &
45.62$^{+0.11}_{-0.11}$ & 45.82 & $-$1.63 & $+$0.14 (1.08) & $<$
4.3\tablenotemark{f} \\ 170100.62$+$641209.0 & 2.53 & 16.0 & $-$30.24
& 17.84 & 32.47 & 6.9$^{+0.5}_{-0.4}$ & 34.4$^{+2.2}_{-2.1}$ &
19.2$^{+1.2}_{-1.2}$ & 45.19$^{+0.03}_{-0.03}$ & 45.40 & $-$1.91 &
$-$0.10 (0.73) & $<$ 1.1\tablenotemark{f} \\ 173352.22$+$540030.5 &
3.36 & 17.0 & $-$29.54 & 6.92 & 32.22 & 8.9$^{+1.8}_{-1.5}$ &
36.1$^{+7.5}_{-6.3}$ & 23.9$^{+5.0}_{-4.1}$ & 45.44$^{+0.08}_{-0.08}$
& 45.56 & $-$1.71 & $+$0.06 (0.48) & 10.0 \\ 212329.46$-$005052.9 &
4.78 & 16.5 & $-$29.38 & 11.21 & 32.13 & 5.6$^{+1.5}_{-1.2}$ &
23.7$^{+6.2}_{-5.0}$ & 11.6$^{+3.0}_{-2.4}$ & 44.82$^{+0.10}_{-0.10}$
& 45.20 & $-$1.91 & $-$0.15 (1.15) & $<$ 0.3 \\ 231324.45$+$003444.5 &
4.03 & 16.4 & $-$29.56 & 13.12 & 32.13 & 2.6$^{+2.6}_{-1.4}$ &
11.4$^{+11.4}_{-6.4}$ & 5.2$^{+5.2}_{-2.9}$ & 44.42$^{+0.30}_{-0.36}$
& 44.62 & $-$2.07 & $-$0.31 (2.35) & $<$ 0.2 \\ \cline{1-14} APM
08279$+$5255 & 4.05 & 15.1 & $-$32.00 & 43.87 & 33.11 &
44.7$^{+0.7}_{-0.7}$ & 142.9$^{+2.3}_{-2.2}$ & 103.9$^{+1.7}_{-1.6}$ &
46.17$^{+0.01}_{-0.01}$ & 46.37 & $-$1.78 & $-$0.16
(1.07)\tablenotemark{g} & $<$ 0.2 \\ HS 1603$+$3820 & 1.32 & 16.1 &
$-$30.05 & 17.94 & 32.41 & 11.3$^{+1.3}_{-1.2}$ & 35.4$^{+4.0}_{-3.7}$
& 18.5$^{+2.1}_{-1.9}$ & 45.11$^{+0.05}_{-0.05}$ & 45.23 & $-$1.91 &
$-$0.11 (0.85) & $<$ 0.2 \\ \tableline \enddata \tablenotetext{a}{Flux
density at rest-frame wavelength 2500~\AA\ in units of 10$^{-27}$ ergs
cm$^{-2}$ s$^{-1}$ Hz$^{-1}$.}  \tablenotetext{b}{Observed count rate
computed in the \hbox{0.5--2~keV} band in units of $10^{-3}$ counts
s$^{-1}$.  The count rates for off-axis sources have been corrected
for vignetting.}  \tablenotetext{c}{Galactic absorption-corrected flux
in the observed \hbox{0.5--2~keV} band in units of $10^{-15}$ erg
cm$^{-2}$ s$^{-1}$.}  \tablenotetext{d}{Flux density at rest-frame
2~keV in units of $10^{-32}$ ergs cm$^{-2}$ s$^{-1}$ Hz$^{-1}$.}
\tablenotetext{e}{The difference between measured and predicted \aox\
($\Delta$\aox), and the significance of that difference ($\sigma$),
based on the S06 \aox--$L_{\nu}(2500~\mbox{\AA})$ relation.}
\tablenotetext{f}{Flux density at an observed-frame frequency of
1.4~GHz taken from the NVSS survey. All other 1.4~GHz flux densities
are from the FIRST survey.}  \tablenotetext{g}{\daox values for
gravitationally lensed objects have been calculated using
lensing-corrected luminosities.}
\label{tab4}
\end{deluxetable}
\clearpage
\end{landscape}

\end{document}